\documentclass[fleqn, usenatbib]{mnras}

\usepackage{newtxtext,newtxmath}
\usepackage{footmisc}
\usepackage[T1]{fontenc}
\usepackage{ae,aecompl}
\usepackage{gensymb}
\usepackage{acronym}
\usepackage{relsize}
\usepackage{graphicx}   
\usepackage{amsmath}    
\usepackage{siunitx}    
    \DeclareSIUnit \parsec {pc}
    \DeclareSIUnit \year {yr}
    \DeclareSIUnit \AU {AU}
    \DeclareSIUnit \erg {erg}
    \DeclareSIUnit \solarmass {M_{\odot}}
    \DeclareSIUnit \angstrom {\text {Å}}
    \DeclareSIUnit \pixel {pix}
    \DeclareSIUnit \jansky {Jy}
\usepackage{dirtytalk} 
\usepackage{url}
\usepackage{enumitem}
\usepackage{subcaption}
\usepackage{hyperref}
\acrodef{HST}[\emph{HST}]{\emph{Hubble Space Telescope}}
\acrodef{ICM}{intra-cluster medium}
\acrodef{SF}{star-forming}
\acrodef{ISM}{interstellar medium}
\acrodef{BCG}{brightest cluster galaxy}
\acrodef{SFR}{star formation rate}
\acrodef{SFH}{star formation history}
\acrodef{sSFR}{specific star formation rate}
\acrodef{AGN}{active galactic nuclei}
\acrodef{DM}{dark matter}
\acrodef{CGM}{circumgalactic medium}
\acrodef{IMF}{initial mass function}
\acrodef{WFC3}{Wide Field Camera 3}
\acrodef{GLASS}{Grism Lens-Amplified Survey from Space}
\acrodef{ESO-NTT}{European Southern Observatory New Technology Telescope}
\acrodef{SN}{supernova}
\acrodef{MUSE}{Multi Unit Spectroscopic Explorer}
\acrodef{MOSFIRE}{Multi-Object Spectrometer For Infra-Red Exploration}
\acrodef{SSP}{simple stellar population}
\acrodef{RPS}{Ram pressure stripping}
\acrodef{FOV}{field-of-view}
\acrodef{WFC3}{Wide-Field Camera 3}
\acrodef{EoR}{Epoch of Reionization}
\acrodef{SED}{spectral energy distribution}
\acrodef{PSF}{point-spread function}
\acrodef{EW}{equivalent width}
\acrodef{DOF}{degrees of freedom}
\acrodef{FWHM}{full width half maximum}
\acrodef{FUV}{far-ultraviolet}
\acrodef{UV}{ultraviolet}
\acrodef{NUV}{near-ultraviolet}
\acrodef{NIR}{near-infrared}
\acrodef{IR}{infrared}
\acrodef{FIR}{far-infrared}
\acrodef{HAEs}{H$\alpha$ emitters}
\acrodef{PDF}{probability distribution function}

\title[Size of $z=6.1$ H$\alpha$-emitters with JWST]{The \emph{JWST} Emission Line Survey (JELS): The sizes and merger fraction of star-forming galaxies during the Epoch of Reionization}

\author[H.M.O. Stephenson et al.]{H. M. O. Stephenson,$^{1}$\thanks{E-mail: h.stephenson@lancaster.ac.uk (HMOS)}
J. P. Stott,$^{1}$\thanks{E-mail: j.p.stott@lancaster.ac.uk (JPS)}
C. A. Pirie,$^{2}$
K. J. Duncan,$^{2}$
D. J. McLeod,$^{2}$
P. N. Best,$^{2}$
M. Brinch,$^{3}$\newauthor
M. Clausen,$^{2}$
R. K. Cochrane,$^{4, 2, 5}$
J. S. Dunlop,$^{2}$
S. R. Flury,$^{2}$
J. E. Geach,$^{6}$
C. L. Hale,$^{7}$
E. Ibar,$^{3, 8}$\newauthor
Zefeng Li,$^{9}$
J. Matthee,$^{10}$
R. J. McLure,$^{2}$
L. Ossa-Fuentes,$^{3}$
A. L. Patrick,$^{2}$
D. Sobral,$^{11, 12}$\newauthor
and A. M. Swinbank$^{9}$
\\
$^{1}$Department of Physics, Lancaster University, Lancaster, LA1 4YB, UK\\
$^{2}$Institute for Astronomy, University of Edinburgh, Royal Observatory, Blackford Hill, Edinburgh, EH9 3HJ, UK\\
$^{3}$Instituto de F\'isica y Astronom\'ia, Universidad de Valpara\'iso, Avda. Gran Breta\~na 1111, Valpara\'iso, Chile\\
$^{4}$Jodrell Bank Centre for Astrophysics, University of Manchester, Oxford Road, Manchester M13 9PL, UK\\
$^{5}$Department of Astronomy, Columbia University, New York, NY 10027, USA\\
$^{6}$Centre for Astrophysics Research, School of Physics, Engineering and Computer Science, University of Hertfordshire, Hatfield, UK\\
$^{7}$Astrophysics, Department of Physics, University of Oxford, Denys Wilkinson Building, Keble Road, Oxford, OX1 3RH, UK\\
$^{8}$Millennium Nucleus for Galaxies (MINGAL), Valparaíso, Chile\\
$^{9}$Centre for Extragalactic Astronomy, Department of Physics, Durham University, South Road, Durham DH1 3LE, UK\\
$^{10}$Institute of Science and Technology Austria (ISTA), Am Campus 1, 3400 Klosterneuburg, Austria\\
$^{11}$Departamento de F\'isica, Faculdade de Ci\`encias, Universidade de Lisboa, Edif\'icio C8, Campo Grande, PT1749-016 Lisbon, Portugal\\
$^{12}$BNP Paribas Corporate \& Institutional Banking, Torre Ocidente Rua Galileu Galilei, 1500-392 Lisbon, Portugal
}

\date{Accepted XXX. Received YYY; in original form ZZZ}

\pubyear{2025}
\begin{document}
\label{firstpage}
\setlength{\abovedisplayskip}{0pt}
\setlength{\belowdisplayskip}{0pt}
\setlength{\parskip}{0pt}
\pagerange{\pageref{firstpage}--\pageref{lastpage}}
\maketitle

\begin{abstract}
We used observations from the \emph{JWST} Emission Line Survey (JELS) to measure the half-light radii ($r_{e}$) of 23 H$\alpha$-emitting star-forming (SF) galaxies at $z=6.1$ in the PRIMER/COSMOS field. Galaxy sizes were measured in \emph{JWST} Near-infrared Camera observations in rest-frame H$\alpha$ (tracing recent star formation) with the F466N and F470N narrowband filters from JELS, and compared against rest-$R$-band, $V$-band (tracing established stellar populations) and near-ultraviolet sizes. We find a size-stellar mass ($r_{e}-M_{*}$) relationship with a slope that is consistent with literature values at lower redshifts, though offset to lower sizes. We observe a large scatter in $r_{e}$ at low stellar mass ($M_{*}<10^{8.4}$\,\si{\solarmass}) which we believe is the result of bursty star formation histories (SFHs) of SF galaxies at the Epoch of Reionization (EoR). We find that the stellar and ionised gas components are similar in size at $z=6.1$. The evidence of already-established stellar components in these H$\alpha$ emitters (HAEs) indicates previous episodes of star formation have occurred. As such, following other JELS studies finding our HAEs are undergoing a current burst of star formation, we believe our results indicate that SF galaxies at the end of the EoR have already experienced a bursty SFH. From our $r_{e}-M_{*}$ relationship, we find $r_{e, \text{F444W}}=0.76\pm0.46$\,\si{\kilo\parsec} for fixed stellar mass $M_{*}=10^{9.25}$\,\si{\solarmass}, which is in agreement with other observations and simulations of star forming galaxies in the literature. We find a close-pair (major) merger fraction of ($f_{\text{maj. merger}}=0.44\pm0.22$) $f_{\text{merger}}=0.43\pm0.11$ for galaxy separations $d\lesssim25$\,\si{\kilo\parsec}, which is in agreement with other $z\approx6$ studies.
\end{abstract}




\section{Introduction}
\label{sec::intro}


The redshift evolution of the basic physical properties of galaxies, such as their size, stellar mass ($M_{*}$), luminosity, and morphology provide vital constraints for models of galaxy formation. Additionally, how these properties change with respect to each other can constrain the evolutionary tracks that galaxies follow. One example is the size distribution of galaxies as a function of their stellar mass. Galaxy size is often measured in units of `effective' or `half-light' radius ($r_{e}$), defined as the radius in which half of a galaxy's light is contained. $r_{e}$ can range from $\approx0.1$\,\si{\kilo\parsec} (e.g. \citealp{Ono2023}) to $\gtrsim10$\,\si{\kilo\parsec} (e.g. \citealp{Kawamata2015,Dullo2017}). The distribution and evolution of $r_{e}$ can be used to infer properties of the host \ac{DM} halo, including its virial radius \citep{Mo1998,Dutton2007,Fu2010}, spin or angular momentum \citep{Bullock2001,Dutton2009a}, and merger history \citep{Naab2009,Ownsworth2014}.

There is strong evidence that galaxy sizes correlate with stellar mass such that higher mass galaxies have a larger $r_{e}$ (the size-mass relationship; $r_{e}-M_{*}$). In their influential work, \citet{Shen2003} studied the $r_{e}$ distributions of $\approx$140 000 galaxies in the Sloan Digital Sky Survey (SDSS; \citealp{York2000,Stoughton2002}) as a function of stellar mass and luminosity and found a clear correlation with both. They also found that both relations are significantly steeper for early-type galaxies than late-type galaxies, the latter of which has a characteristic stellar mass in the local Universe ($M_{*, 0} = 10^{10.6}$\,\si{\solarmass}) above which the slope steepens but remains shallower than that of early-type galaxies. \citet{vanderWel2014a} extended the analysis of the mass-size relation out to higher redshifts, covering $0 \lesssim z \lesssim 3$ by making use of \ac{HST} data from the 3D-HST survey \citep{Brammer2012} and the Cosmic Assembly Near-infrared Deep Extragalactic Legacy Survey (CANDELS; \citealp{Grogin2011,Koekemoer2011}). The results of \citet{vanderWel2014a} are in qualitative agreement with \citet{Shen2003} such that the slope of the $r_{e}-M_{*}$ relation is shallower for late-type galaxies, though they see a flattening in the slope of early-type galaxies at $M_{*} \lesssim 10^{10}$\,\si{\solarmass}. Many studies have measured $r_{e}-M_{*}$ relations at a range of redshifts ($0<z<5$; \citealp{Daddi2005,Trujillo2006,Trujillo2007,Stott2011,Stott2013a,Lange2015,Paulino-Afonso2017,Faisst2017,Mowla2019a,Mowla2019b}). More recently, studies have utilised the high resolution of \emph{JWST} \citep{Gardner2006,Rigby2023} in the \ac{NIR} to study the structural properties of galaxies in the rest-frame optical at $z\gtrsim3$ including sizes \citep{Suess2022,Ono2023,Ormerod2023,Allen2025,Ji2024a,Martorano2024,Varadaraj2024,Ward2024,Miller2025,Westcott2025}, resolved star formation \citep{Ji2024b,Li2024,Matharu2024,Morishita2024} and morphology \citep{Ito2024,Ono2024,Ono2025,Vega-Ferrero2024}.

As mentioned above, there has been a clear redshift evolution of $r_{e}$ and the $r_{e}-M_{*}$ relationship. \citet{vanderWel2014a} found that their $r_{e}$ measurements at $z\sim0$ are consistent with those of SDSS galaxies in \citet{Shen2003} and \citet{Guo2009} (after accounting for systematic differences in their respective methods; see \citealp{vanderWel2014a}), but observed an increasing offset from the local $r_{e}-M_{*}$ relation for higher redshift galaxies. The power-law that they use to describe the redshift evolution of galaxy sizes is of the form

\begin{equation}
\frac{r_{e}}{\text{kpc}} = B(1+z)^{\beta},
\label{eq::sizezpower}
\end{equation}

\vspace{5pt}
\noindent where $B$ is the intercept and $\beta$ is the power-law slope. Equation \ref{eq::sizezpower} is used as standard in the literature to describe the redshift evolution of $r_{e}$, typically for some characteristic stellar mass \citep{Stott2013a,Shibuya2015,Paulino-Afonso2017,Cutler2022,Sun2024,vanderWel2024}. Specifically, \citet{vanderWel2014a} find that for a fixed stellar mass of $10^{10.75}$\,\si{\solarmass}, the sizes of late- and early-type galaxies evolve as $r_{e} \propto (1+z)^{-0.72}$ and $r_{e} \propto (1+z)^{-1.24}$ respectively. These results reflect an $\approx3.7$\,\si{\kilo\parsec} ($\approx55\%$) and $\approx3.2$\,\si{\kilo\parsec} ($\approx74\%$) decrease in $r_{e}$, respectively, between $z<0.5$ and $z=2.5-3$. Using \emph{JWST} data from The Cosmic Evolution Early Release Survey (CEERS; \citealp{Finkelstein2023,Bagley2023}) and \emph{HST} data from CANDELS, \citet{Ward2024} found that the average $r_{e}$ decreased further out to $z=5.5$. Their power law slope of $\beta = -0.67\pm0.07$ for a characteristic mass of $\approx10^{10.7}$\,\si{\solarmass} is also consistent with that of \citet{vanderWel2014a}. It is crucial to extend analysis of galaxy sizes to even higher redshifts, and for homogeneously selected populations, in order to constrain this power-law further and to shed new light on the first era of galaxy formation.

In order to disentangle the contributions of active star formation from the longer-term, in-situ star formation history (SFH) on the size evolution discussed above, one must use \ac{SFR} indicators that are distinct to both, as well as high spatial resolution images to resolve active \ac{SF} regions. The H$\alpha$ (6563\,\si{\angstrom}) emission line is one of the most frequently used \ac{SFR} indicators for recent star formation in a galaxy \citep{Kennicutt1998,Erb2006b,Sobral2013a,Terao2022,Covelo-Paz2025}. This is because it traces the ionised gas emission resulting from the recombination of hydrogen surrounding the most massive stars \citep{Hao2011a,Murphy2011}, which typically only live for $\lesssim10$\,\si{\mega\year} \citep{Ekstrom2012}. The rest-frame \ac{UV} or \ac{NUV} continuum of a galaxy can also be used to trace star formation, but on longer timescales ($\sim10-200$\,\si{\mega\year}) as it traces the photons emitted directly from the photospheres of stars upwards of several solar masses. For a comprehensive review of \ac{SFR} indicators and the populations they trace, see \citeauthor{Kennicutt2012} (\citeyear{Kennicutt2012}; also \citealp{Madau2014,Sanchez2020,Schinnerer2024}). Typically, the \ac{UV} continuum has been used to infer the \ac{SFR} of galaxies at $z\gtrsim2$ as its wavelength gets redshifted into optical bands \citep{Wyder2005,Bouwens2012a,Bouwens2012b,Bouwens2015a,Oesch2018,Harikane2023}, whereas H$\alpha$ is shifted further into \ac{IR} bands which are more difficult to observe with ground based instruments. However, \ac{UV} measurements are extremely sensitive to dust attenuation (e.g. \citealp{Calzetti1994,Dunlop2017,Bouwens2020,Traina2024}), and there is evidence that the effects of dust continue to impact observations out to the Epoch of Reionization (EoR; \citealp{Gruppioni2020,Bowler2022,Algera2023,Zavala2023})\acused{EoR}. Moreover, while there exist dust attenuation calibrations that aim to correct for these issues (e.g. \citealp{Calzetti2000,Salim2007}), these depend heavily on assumptions of the \ac{UV} continuum slope, intrinsic colours and the choice of dust extinction curve which may impact the measured properties of these selected objects \citep{Walcher2011,ArrabalHaro2023}.

The benefit of the H$\alpha$ emission line is that it is less affected by dust attenuation, which relieves these issues (though we note that the extinction in the nebula regions is more uncertain at high-$z$; \citealp{Reddy2020,Reddy2025}). With \emph{JWST} able to access H$\alpha$ at $z\gtrsim2.5$, combined with its high resolution imaging, we are now able to probe the physical properties of samples of \ac{SF} galaxies at earlier times than before. Some studies are already showcasing the ability to probe star formation using H$\alpha$ out to $z\sim3.7-6.5$ using grism spectroscopy to select on H$\alpha$ \citep{Matharu2023,Nelson2024,Covelo-Paz2025}. Another approach is to use narrowband (NB) imaging selection on H$\alpha$ to look for \ac{SF} galaxies, as this also provides a selection based on the strength of the emission line. A key advantage of NB imaging is that it mitigates the selection effects that can arise from slitless spectroscopy (such as source blending, or preferentially strong Lyman breaks) and, when combined with broadband (BB) photometry, provides a much narrower redshift range that sources can lie within. Additionally, it also provides a direct image without needing to reconstruct one from the slitless spectroscopy. This method was notably used by the Hi-Z Emission Line Survey (HiZELS; \citealp{Geach2008,Sobral2009,Sobral2013a}) which utilised the Wide Field Camera (WFCAM; \citealp{Casali2007}) on the United Kingdom Infrared Telescope (UKRIT) to significantly expand the volume of previous narrowband imaging surveys (e.g. \citealp{Thompson1996,Moorwood2000}). As a result of the narrowband imaging selection of HiZELS, many studies were able to measure the properties of homogeneously-selected \ac{HAEs} out to $z=2.23$ \citep{Sobral2010,Garn2010a,Swinbank2012a,Swinbank2012b,Sobral2012,Sobral2013a,Stott2013b,Oteo2015,Sobral2016a,Cochrane2017,Cochrane2018,Cheng2020b,Cochrane2021}, including $r_{e}$ measurements \citep{Stott2013a,Paulino-Afonso2017,Naufal2023}.

In this paper, we use data from the \emph{JWST} Emission Line Survey (JELS; GO \#2321; PI: Philip Best; see \citealp{Duncan2025a,Pirie2025a}) to probe the $r_{e}$ properties of 23 $z=6.1$ \ac{SF} galaxies in the first H$\alpha$-selected sample of \ac{HAEs} from NB imaging at the \ac{EoR}. We combine the JELS observations with anicillary multi-wavelength data from the \emph{JWST} Cycle 1 Observer Treasury Program `Public Release IMaging for Extragalactic Research' survey (PRIMER; PI Dunlop, GO \#1837)\footnote{\label{fn::primer}\url{https://primer-jwst.github.io}}. We primarily use \emph{JWST}/NIRCam long-wavelength (LW) channel observations in JELS F466N, JELS F470N NB and PRIMER F444W BB filters to study the rest-frame $R$-band S\'ersic light profiles \citep{Sersic1963,Sersic1968} of \ac{HAEs} at the \ac{EoR}, taking advantage of the image resolution \emph{JWST} provides at $\lambda\approx3.8-5$\,\si{\micro\meter}. This allows us to probe both active star formation and older stellar populations at $z=6.1$. In addition to rest-$R$-band sizes, we also measure the light profiles in PRIMER F277W (rest-\ac{NUV}) and PRIMER F356W (rest-$V$-band). Our measured $r_{e}$ values are then compared to both observations and simulations. We also measure the size ratio of the \ac{SF} component (traced by the ionised gas from the H$\alpha$ emission) and the stellar continuum to infer how \ac{EoR} \ac{HAEs} have evolved over the preceding 1\,\si{\giga\year}.

This paper is arranged as follows. In Section \ref{sec::data}, we summarise the JELS survey and photometric catalogue. We explain how we determine our final sample of \ac{HAEs} in Section \ref{subsec::sampleselection}. In Section \ref{sec::sersicmodel}, we describe the methods used to fit galaxy sizes in different wavebands. We outline key results in Section \ref{sec::results}, and discuss their implications in Section \ref{sec::discussion}. We summarise our conclusions in Section \ref{sec::conclusions}.

A standard $\Lambda$CDM cosmology model is assumed with values $\Omega_{\Lambda} = 0.7$, $\Omega_{m} = 0.3$ and $H_{0} = 70$\,\si{\kilo\meter\per\second\per\mega\parsec}. Any magnitudes stated are presented using the AB system \citep{Oke1983}. All results and comparisons to the literature in this paper assume a \citet{Chabrier2003} \ac{IMF}.


\section{JELS Data}
\label{sec::data}

The JELS survey is described in full by \citet{Duncan2025a} and \citet{Pirie2025a}. Here, we will summarise the key details that are relevant for our work.

The primary goal of JELS is to provide a homogeneously-selected catalogue of H$\alpha$-selected galaxies at the \ac{EoR} from the COSMOS field \citep{Scoville2007a,Scoville2007b}. In this context, \say{homoegenously-selected} refers to the fact galaxies are identified solely based on their H$\alpha$ emission-line strength, providing a uniform tracer of star formation and avoiding biases introduced by continuum- or colour-based selection methods. This is, in effect, a selection on \ac{SFR}, though we note here the catalogue described in this Section is complete in stellar mass to $\approx10^{8.2}$\,\si{\solarmass}. This selection is achieved by employing the F466N and F470N NB filters in the \emph{JWST}/NIRCam LW channels, with pivot wavelengths of $\lambda_{\text{pivot}} = 4.654$\,\si{\micro\metre} (effective width $W_{\text{eff}} = 0.0535$\,\si{\micro\metre}) and $\lambda_{\text{pivot}} = 4.707$\,\si{\micro\metre} ($W_{\text{eff}} = 0.0510$\,\si{\micro\metre}) respectively. These two filters centre on $z\approx6.09$ and $z\approx6.17$ for the H$\alpha$ emission line which allows for line emitters to be selected through difference imaging in a selection volume of $\sim2.4\times10^{4}$\,\si{\mega\parsec\cubed}. In addition, the NB observations from JELS are designed to overlap with multi-wavelength observations from CANDELS \citep{Grogin2011,Koekemoer2011,Brammer2012,Teplitz2018} and, more crucially, with the \emph{JWST} Cycle 1 Observer Treasury Program PRIMER survey (PI Dunlop, GO \#1837)\footref{fn::primer}. Specifically, LW BB observations in the F444W filter ($\lambda_{\text{pivot}} = 4.4043$\,\si{\micro\metre}; $W_{\text{eff}} = 1.0676$\,\si{\micro\metre}) from PRIMER - with a wavelength range that covers F466N and F470N - allow for this NB excess selection at $\sim4.7$\,\si{\micro\metre}. In this work, we also make use of PRIMER observations in F277W ($\lambda_{\text{pivot}} = 2.7617$\,\si{\micro\metre}; $W_{\text{eff}} = 0.6615$\,\si{\micro\metre}) and F356W ($\lambda_{\text{pivot}} = 3.5684$\,\si{\micro\metre}; $W_{\text{eff}} = 0.7239$\,\si{\micro\metre}) for rest-\ac{NUV} and rest-$V$-band measurements respectively. The resulting deep, multi-wavelength coverage has enabled robust photometry for \ac{SED} fitting to constrain the SFH of the \ac{HAEs}. The \ac{SED} fitting for our \ac{HAEs} was performed by \citet{Pirie2025a} using the \texttt{BAGPIPES} spectral fitting code \citep{Carnall2018}. They use \texttt{BPASS} \citep{Eldridge2017,Stanway2018} for their stellar population synthesis model, and the \texttt{CLOUDY} photoionisation code \citep{Ferland2017} for nebular emission line computation. They assume a \citet{Salim2018} dust attenuation model and the continuity non-parametric SFH model from \citet{Leja2019}. We refer the reader to Table 8 of \citet{Pirie2025a} for details on the models and priors.

Overall, the JELS survey has continuous coverage over $\sim63$\,arcmin$^{2}$ area of the COSMOS field (see Figure 3 of \citealp{Duncan2025a}) with the final images homogenised to a common \ac{PSF} with a resolution of $0.03\times0.03$\,arcsec$^{2}$ per pixel.

We note here that the science conducted in this paper is based on the initial versions of the JELS imaging products, referred to as v0.8 images, also used in \citet{Pirie2025a}. The newer v1.0 images incorporate re-observations taken in November 2024 to better mitigate scattered light issues and use an updated version of the \emph{JWST} pipeline for image reduction. We refer the reader to Appendix A of \citet{Duncan2025a} for details on the differences between the v0.8 images used here and the updated v1.0 images. Throughout this work, all references to source IDs are referring to their JELS v0.8 catalogue value and may not correspond to subsequent revisions of the catalogue.

\subsection{Sample Selection}
\label{subsec::sampleselection}

To analyse the structural properties of \ac{SF} galaxies at the \ac{EoR}, we derive our own sample from the multi-wavelength JELS v0.8 catalogues that are described in detail by \citet{Pirie2025a}, specifically their Sections 3 and 4 where the method for identifying NB excess emitters from multi-wavelength detections is outlined. Briefly, their selection is based on significant NB excess relative to either the overlapping PRIMER F444W BB or neighbouring NB filter, combined with photo-$z$ cuts around $z\sim6.1$, with additional visual inspection to remove residual contamination. Following this, they finalise a catalogue of 35 \ac{HAEs} (30 F466N sources; 5 F470N sources). We refer to the \citet{Pirie2025a} catalogue as the `parent' catalogue in this paper, and it is from this that we determine the sample used for the analysis in this work.

In order to accurately measure $r_{e}$, we need a sample that we are able to reliably measure the S\'ersic profiles of. Therefore, we removed sources from the parent catalogue of \citet{Pirie2025a} that had some additional complications which made them difficult/impossible to reliably model. Here, we will outline which sources were removed, reiterating that the IDs refer to the JELS v0.8 catalogue. Sources 2768 and 7810 (both F466N selected) were removed because light profile fits in both NB and F444W images strongly preferred a \ac{PSF} model over a S\'ersic model (based on the $\chi^{2}$ outputs of the \texttt{GALFIT} model; see Section \ref{subsec::galfitfitting}), indicative of a point source dominating any galactic emission; 12164 (F466N) was removed because a \ac{PSF} model was strongly preferred in the F444W image (see Section \ref{subsec::galfitfitting}) meaning point source emission is likely dominating; 4453 and 4457 (both F466N) were removed as they appear to be a merging system (see Section \ref{subsec::merger_rates}) that is too faint to individually model in the F444W image; 10983 (F466N) was removed because, despite being an isolated source in the NB image, it appears to be part of a three-way merging system in F444W which made it difficult to isolate when modelling. Finally, a number of \ac{HAEs} were too faint in F444W to accurately model. From F466N detections, 5629 ($\approx 28.7$\,mag), 6501 ($\approx 28.8$\,mag), 7147 ($\approx 29.3$\,mag) and 9123 ($\approx 28.7$\,mag) were removed, as well as 8033 ($\approx 30.1$\,mag) and 15619 ($\approx 28.8$\,mag) from F470N detections. See Figure A1 of \citet{Pirie2025a} for multi-wavelength imaging of all excess sources, including those we do not include in our final \ac{HAEs} sample.

Following these removals, we obtain a sample of 23 H$\alpha$-emitting \ac{SF} galaxies at $z=6.1$ (20 detected in F466N; 3 detected in F470N). These galaxies have a stellar mass range of $M_{*} = 10^{8.06-9.28}$\,\si{\solarmass} ($M_{*,\text{median}} = 10^{8.30}$\,\si{\solarmass}), and \ac{SFR} range of $\text{SFR}_{\text{H}\alpha} = 1.03 - 14.22$\,\si{\solarmass\per\year} (median $\text{SFR}_{\text{H}\alpha,\text{median}} = 2.73$\,\si{\solarmass\per\year}). Stellar masses and \ac{SFR}s are derived in \citet{Pirie2025a}, where the latter is determined using the H$\alpha$ \ac{SFR} relation in \citet{Theios2019}. We note \ac{SFR} values derived from H$\alpha$ are consistent with the $10$\,\si{\mega\year} \ac{SFR} output from \ac{SED} fitting. All stellar mass and \ac{SFR} values in \citet{Pirie2025a} are converted here to a \citet{Chabrier2003} \ac{IMF}.

The parent catalogue is representative of \ac{HAEs} at $z=6.1$ in terms of H$\alpha$ \ac{EW} (see Figure 11 of \citealp{Pirie2025a}), with an \ac{EW} limit well below the \ac{SF} galaxy population at this redshift (see \citealp{Endsley2024}). Our final sample of 23 \ac{HAEs} is similarly representative in \ac{EW}, and we note that our final sample is also complete in stellar mass down to $\approx10^{8.2}$\,\si{\solarmass}.


\section{S\'ersic Modelling}
\label{sec::sersicmodel}

To measure the S\'ersic light profiles \citep{Sersic1963,Sersic1968} of our sources, we use \texttt{GALFIT}\footnote{\url{https://users.obs.carnegiescience.edu/peng/work/galfit/galfit.html}} version 3.0.7d4 \citep{Peng2002,Peng2010a} - a non-linear least-squares fitting algorithm designed for 2D parametric galaxy fitting. \texttt{GALFIT} uses a Levenberg-Marquardt algorithm to find an optimum solution when modelling the light profile of a source for a given input image. This best-fitting solution is determined via a reduced chi-squared ($\chi^{2}_{\nu}$) method, whereby \texttt{GALFIT} iterates over a large number of possible models by adjusting the input parameters until $\chi^{2}$ is minimised. $\chi^{2}_{\nu}$ describes the goodness-of-fit of the output model and is determined by

\begin{equation}
    \chi^{2}_{\nu} = \frac{1}{N_{\text{DOF}}}\sum_{x=1}^{nx}\sum_{y=1}^{ny} \frac{\left(f_\text{data}(x,y) - f_{\text{model}}(x,y)\right)^{2}}{\sigma(x,y)^{2}},
\label{eq::chisquared}
\end{equation}

\vspace{5pt}
\noindent where $N_{\text{DOF}}$ is the number of \ac{DOF}, $f_{\text{data}}(x,y)$ represents the data image supplied to \texttt{GALFIT}, $f_{\text{model}}(x,y)$ represents the model image that \texttt{GALFIT} outputs following the least-squares fitting and $\sigma(x,y)$ represents the sigma image fed to \texttt{GALFIT}. The sigma image is the relative error of the flux at each position $(x,y)$ within the data image. This is summed over all $x$ and $y$ pixels, where $nx$ and $ny$ are the $x$ and $y$ dimensions, respectively, of the data and model images.

We fit a S\'ersic model of the form

\begin{equation}
    \Sigma(r) = \Sigma_{e}\exp\Biggl\{-b_{n} \left[\left(\frac{r}{r_{e}}\right)^{1/n} - 1\right]\Biggr\},
\label{eq::sersiceq}
\end{equation}
 
\vspace{5pt}
\noindent where $\Sigma(r)$ is the pixel surface brightness at radius $r$ from the centre of a source, $r_{e}$ is the half-light radius of the source and $\Sigma_{e}$ is the pixel surface brightness at $r_{e}$. The S\'ersic index of the model, $n$, determines the overall shape of the light profile and $b_{n}$ is a dimensionless scale factor that is dependent on $n$ (see \citealp{Ciotti1999} for full explanation and asymptotic expansion). In general, late-type galaxies follow a S\'ersic light profile with $n\lesssim2.5$ (shallow inner profile which truncates more sharply at large $r$; \citealp{Sersic1968,Kelvin2012}) and early-type galaxies follow a profile where $n\gtrsim2.5$ (sharply decreasing inner profile with $r$ but extended wing at large $r$; \citealp{Caon1993}). The most commonly used S\'ersic indices are $n = 1$ which describes a purely exponential profile suitable for galactic disks \citep{Freeman1970}, and $n = 4$ which gives a de Vaucouleurs profile \citep{deVaucouleurs1948} suitable for bright elliptical galaxies.


\subsection{Fitting with \texttt{GALFIT}}
\label{subsec::galfitfitting}

\begin{figure*}
    \centering
    \includegraphics[trim={0cm 0 0 0}, width=1\columnwidth]{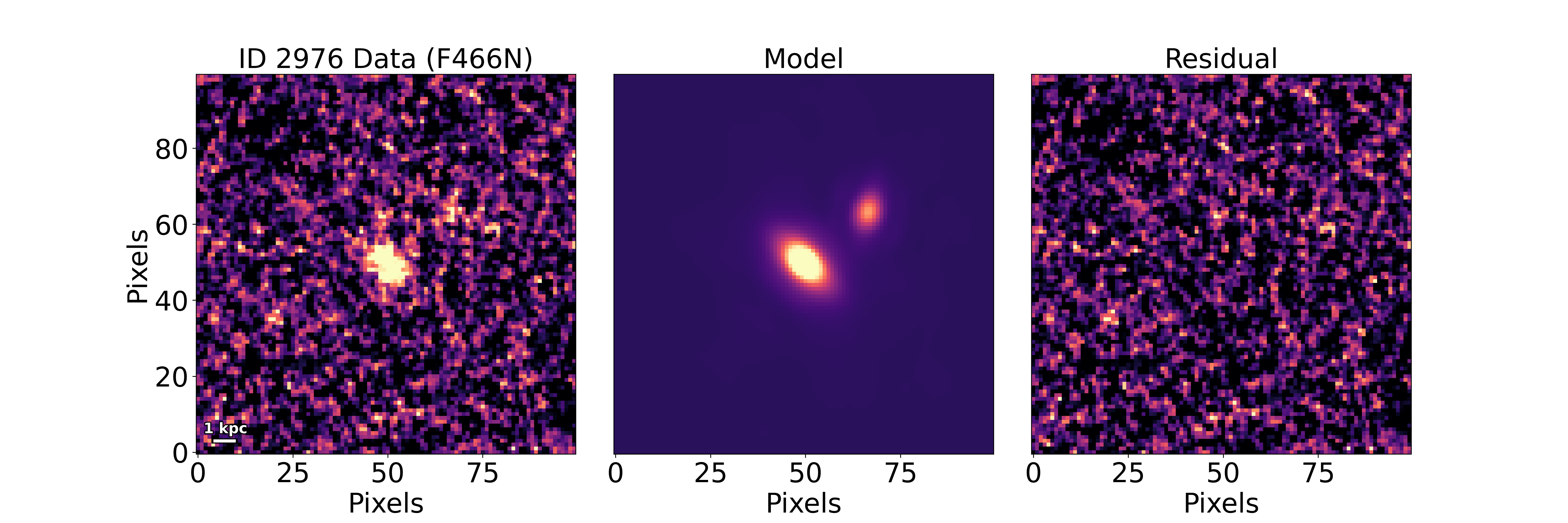}
    \includegraphics[trim={0cm 0 0 0},width=1\columnwidth]{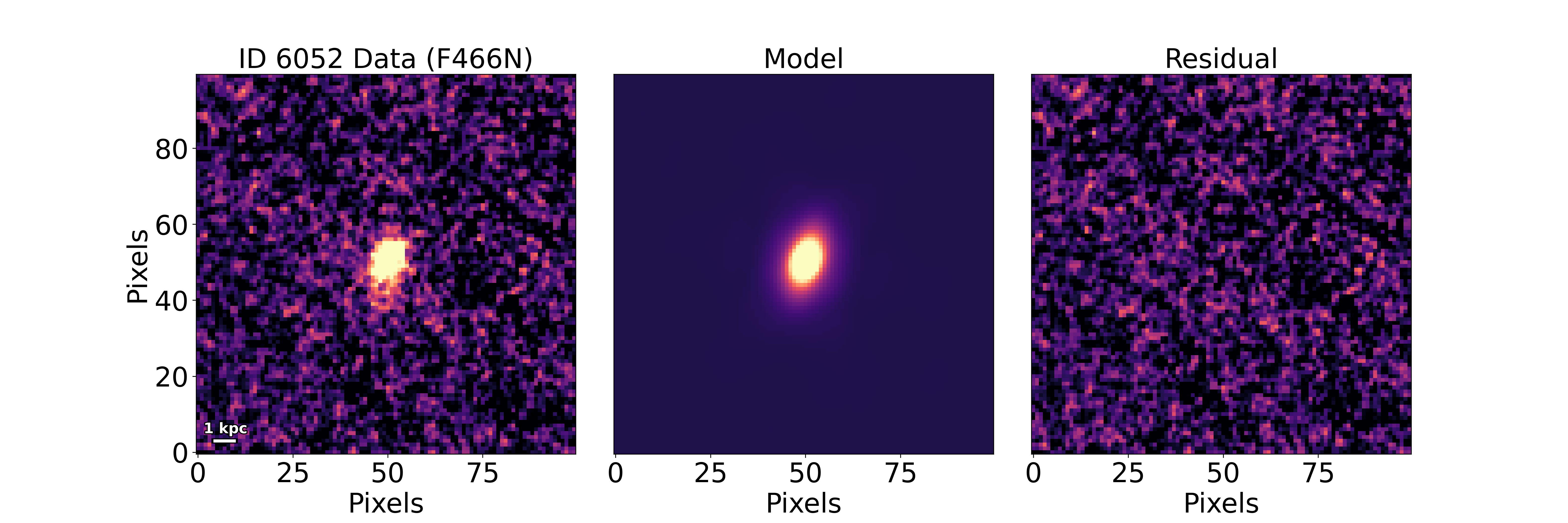}
    \includegraphics[trim={0cm 0 0 0},width=1\columnwidth]{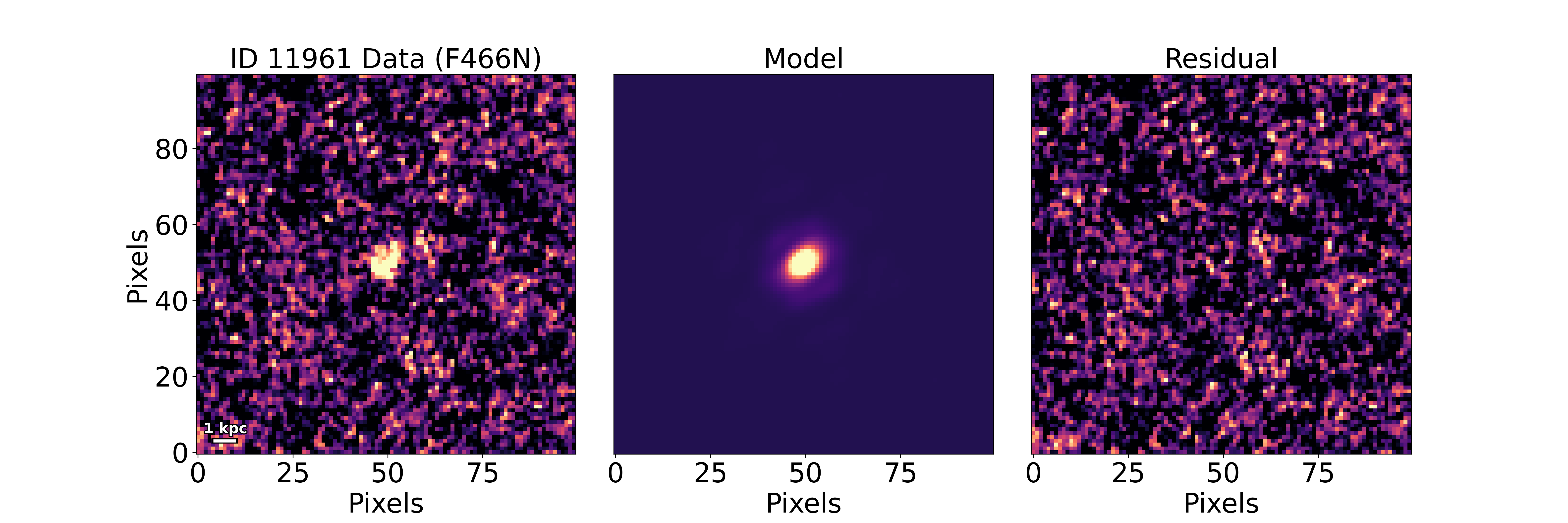}
    \includegraphics[trim={0cm 0 0 0},width=1\columnwidth]{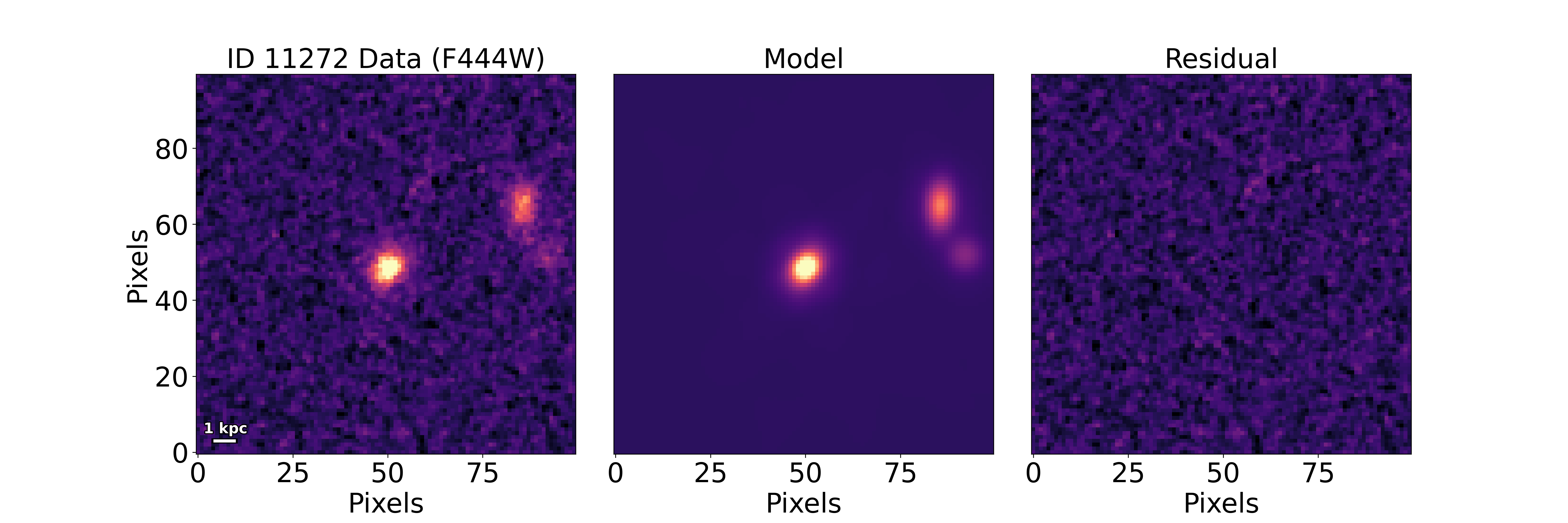}
    \includegraphics[trim={0cm 0 0 0},width=1\columnwidth]{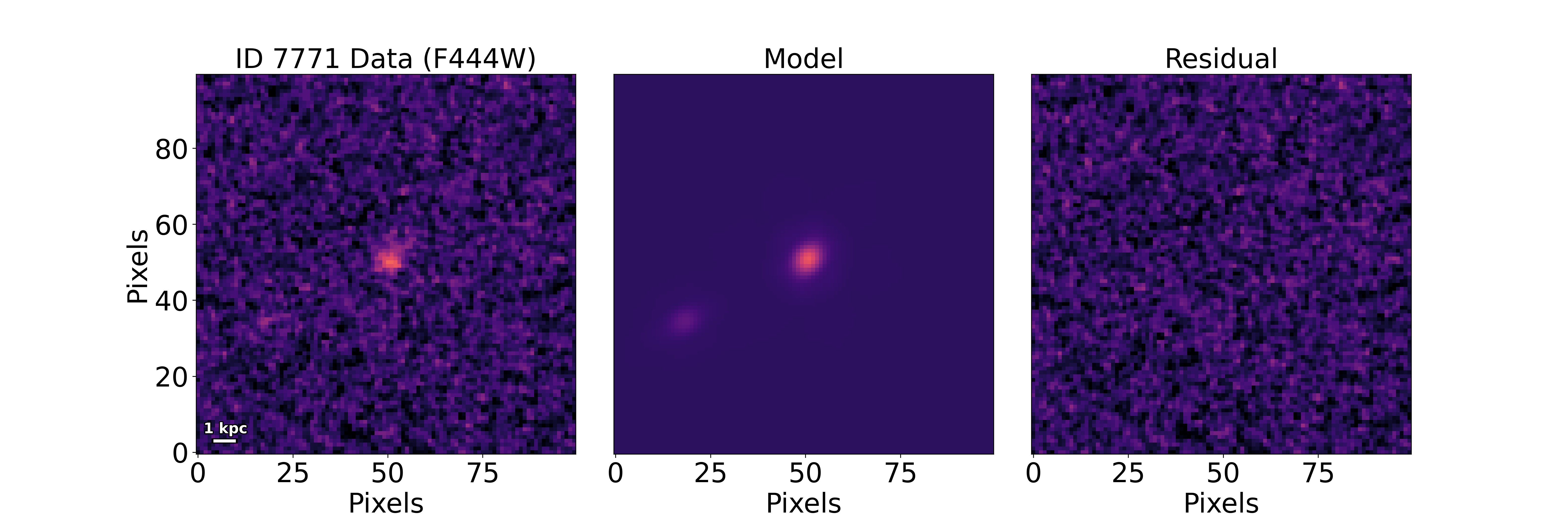}
    \includegraphics[trim={0cm 0 0 0},width=1\columnwidth]{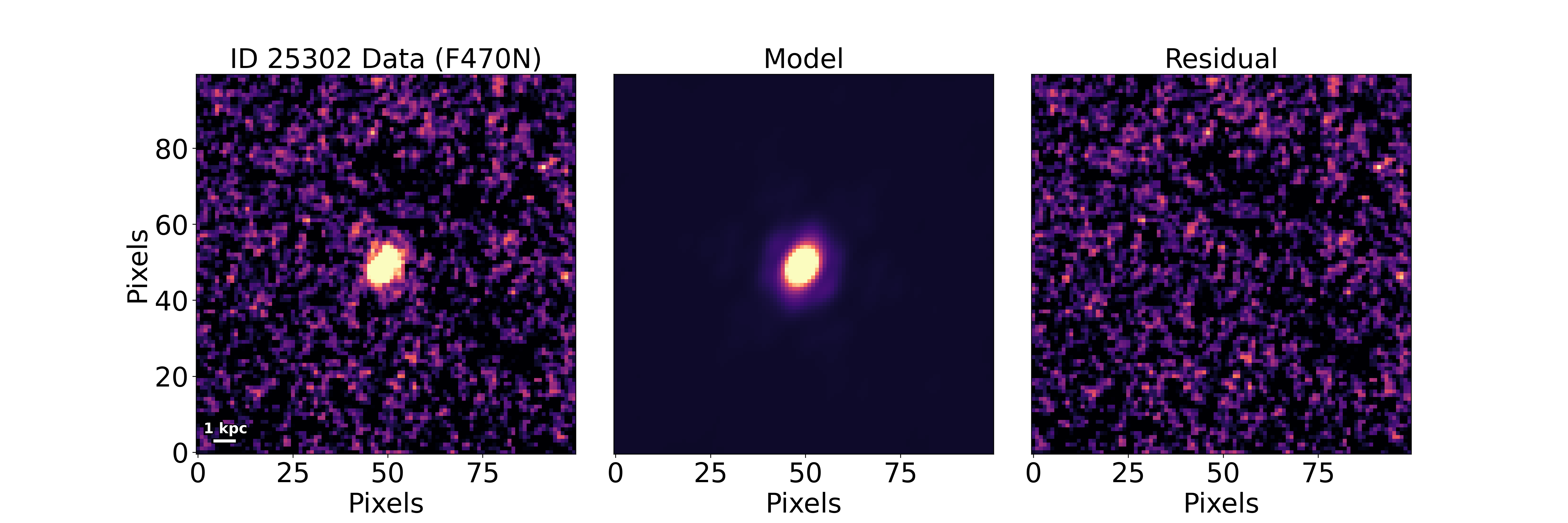}
    \caption{Example models for six \ac{HAEs} in our sample. The left panels are the rest-frame $R$-band observations, with the NIRCam filter used for the observation indicated in the title of the panel. The middle panels are the fixed $n=1$ S\'ersic models of the selected object from \texttt{GALFIT}. The right panels are the residual emission once the modelled galaxy is removed from the observed image. Each panel is a $3\times3$\,arcsec$^{2}$ ($\approx 17\times17$\,\si{\square\kilo\parsec}) cutout centred on the detected galaxy.}
    \label{fig::examplemodels}
\end{figure*}

For our \ac{HAEs}, we fit a single-component S\'ersic profile in all the images we model. We particularly focus on results from the detected JELS F466N and F470N images and the corresponding PRIMER F444W image (equivalent to rest-$R$-band). This is to measure $r_{e}$ of both the H$\alpha$-selected \ac{SF} component and the emission from the stellar population respectively. However, we also follow the procedures in this Section for the PRIMER F277W and PRIMER F356W images for rest-$NUV$ and rest-$V$-band sizes respectively. From this, we can then assess how our results at $z=6.1$ compare to sizes in the literature, including other $r_{e}-M_{*}$ relations. We can also directly compare the NB and F444W sizes to draw conclusions about how \ac{SF} galaxies are evolving at the \ac{EoR} (see Section \ref{subsec::bbtonbsizesec}). The steps we take to model our sources are as follows:

\begin{enumerate}[label=\roman*), leftmargin=*]

    \item First, we create $100\times100$\,pixel$^{2}$ ($3\times3$\,arcsec$^{2}$; $\sim17\times17$\,\si{\square\kilo\parsec}) cutouts of each source centred on the corresponding right ascension (R.A.) and declination (Dec.) of the {\sc SExtractor} \citep{Bertin1996} source coordinates in a given band. We do the same for the corresponding \say{weights} map which is used to create the sigma image to be fed to \texttt{GALFIT}. This weights map has pixel values equal to $1/(\sigma(x,y)^{2})$ so, accordingly, these values are converted such that the sigma image pixel values are $\sigma(x,y)$. Such a relatively large area for the cutout image was decided in order for \texttt{GALFIT} to measure the sky background and confidently capture the wings of the \ac{PSF} (see below). Any additional sources in the cutout, identified via {\sc SExtractor}, are modelled separately so that their light is not accounted for in the selected source model.

    \item As with all telescopes, \emph{JWST} images have an intrinsic \ac{PSF} that must be accounted for \citep{Perrin2014,Rigby2023}. We chose to use empirical $100\times100$\,pixel$^{2}$ \ac{PSF}s described in detail by \citet{Pirie2025a}. In summary, these empirical, filter-dependent \ac{PSF}s were generated by stacking bright and unsaturated stars in a given filter via a boostrapping method. These \ac{PSF}s from \citet{Pirie2025a} are comparable to simulated \ac{PSF}s\footnote{\url{https://jwst-docs.stsci.edu/jwst-near-infrared-camera/nircam-performance/nircam-point-spread-functions}} generated by {\sc WebbPSF}\footnote{\url{https://www.stsci.edu/jwst/science-planning/proposal-planning-toolbox/psf-simulation-tool}} \citep{Perrin2014}. The choice of \ac{PSF} makes no significant difference to our results.

    \item \texttt{GALFIT} requires a set of initial estimates to be provided for each of the fitted parameters. These are the centroid $x$ and $y$ coordinates of the source in pixel units; the integrated apparent magnitude in the chosen filter; $r_{e}$ in pixel units; the S\'ersic index $n$; the semi-minor axis over semi-major axis radius ratio (axis ratio $b/a = q$, where $q = 1$ for a circle and $q<1$ for an ellipse); and the position angle ($\theta_{\text{pa}}$) of the major axis on the sky in degrees measured anti-clockwise from North. Similar to previous studies, we use the {\sc SExtractor} outputs for each of these parameters as our initial guess (e.g. \citealp{vanderWel2012,Mowla2019b,Kartaltepe2023,Ormerod2023,Westcott2025}), with the exception of the integrated magnitude as we used magnitudes derived from flux inside a $0.6$ arcsecond-diameter aperture centred on each source to be consistent with \citet{Pirie2025a}.

    \item The nature of these observations means that these distant galaxies tend to be very small on the image, causing \texttt{GALFIT} to often get stuck in local minima that produce unrealistic output parameters. Additionally, \texttt{GALFIT} will fail if it cannot produce a physical solution \citep{Peng2010a}. To avoid these outcomes, we apply constraints to bound each parameter to be between certain values. These constraints are as follows: the centroid coordinates are allowed to vary $\pm$\,5 pixels from the input values in both $x$ and $y$; the integrated magnitude is allowed to vary $\pm$\,$2$\,mag from the input value; $r_{e}$ is constrained to $0.1\leq r_{e} \leq 100$\,pixels; the axis ratio is constrained to $0.2 \leq q \leq 1$; and $\theta_{\text{pa}}$ is allowed to vary $\pm$\,20$\degree$ from the input values. Each of these constraints are applied to all images. For the S\'ersic index $n$, we initially took two approaches. When modelling in the NB images (F466N and F470N), $n$ is fixed at $n = 1$ since the light profile of any ionised gas emission is expected to be disk-like (e.g. \citealp{Nelson2013}), and our size measurements are consistent regardless of a fixed or free S\'ersic index fit (see right panel of Figure \ref{fig::freevsfixedsizes}). \ac{SF} galaxies at high-$z$ that appear more prolate or oblate in shape have also been shown to have S\'ersic indices of $n\sim1$ (e.g. \citealp{Pandya2024}). For the BB images (F277W, F356W and F444W), we produced two sets of results. One in which $n$ is again fixed at $n = 1$, and another set where we allowed $n$ to take values $0.2 \leq n \leq 8$. The measurements of $r_{e}$ in the BB images using both methods are in agreement (see left panel of Figure \ref{fig::freevsfixedsizes} for results in F444W), and so we chose to fix $n = 1$ for all of our models for ease of interpretation. When the S\'ersic index was left as a free parameter, we find $n_{\text{F444W, median}}\sim1.5$.

    \item It is well documented that \texttt{GALFIT} underestimates the uncertainties of each outputted parameter \citep{Haussler2007,Haussler2013}. Recently, \citet{Ward2024} addressed this by following steps from \citet{vanderWel2012} to recalculate the uncertainty on $r_{e}$ compared to the reported value from \texttt{GALFIT}. They found that their new relative $r_{e}$ errors for their \emph{JWST} images were $\lesssim15\%$, similar to those reported in other studies \citep{vanderWel2012,Nedkova2021}. In light of their findings, we set our uncertainties in $r_{e}$ to be at least $25\%$ of the \texttt{GALFIT} output to be conservative. This is an average factor increase in uncertainty of $\sim3.6$ from the \texttt{GALFIT} output. We refer the reader to Section \ref{subsubsec:galfit_uncertainties} for detailed analysis of $r_{e}$ recovery in \texttt{GALFIT} from known values.

\end{enumerate}

We show six examples of our \texttt{GALFIT} models in Figure \ref{fig::examplemodels}. The sensitivity of \texttt{GALFIT} is alleviated by the constraints described above but following visual inspection, the input centroid coordinates had to be manually adjusted for a small number of models to be closer to the observed centre of the galaxy. None of the adjustments left the {\sc SExtractor} source coordinates outside of the range of estimates \texttt{GALFIT} could take (i.e. manual input was never more than 5 pixels from the initial input) but were necessary adjustments for \texttt{GALFIT} to avoid unphysical local minima in its solution, which we define as being any solution that has an output $r_{e} = 0.1$\,pixels or $r_{e} = 100$\,pixels.

We note that there are potential degeneracies with modelling S\'ersic light profiles with \texttt{GALFIT}, particularly that $r_{e}$ and the S\'ersic index $n$ may be not be independent (see \citealp{Graham2005} for a detailed discussion). This could be a particularly pressing issue for our fixed $n=1$ sizes, though lack of posterior distribution outputs from \texttt{GALFIT} make this difficult to quantify directly \citep{Peng2010a,Haussler2007,Haussler2013}. However, we show in Appendix \ref{sec::appendix} that our sizes are consistent when using either a free S\'ersic or fixed $n=1$ suggesting that the possible limitations of \texttt{GALFIT} do not significantly impact our results.


\subsubsection{Recovering Known Sizes with \texttt{GALFIT}}
\label{subsubsec:galfit_uncertainties}

\begin{figure}
        \centering
        \includegraphics[width=\columnwidth, trim=0 0 0 0, clip=true]{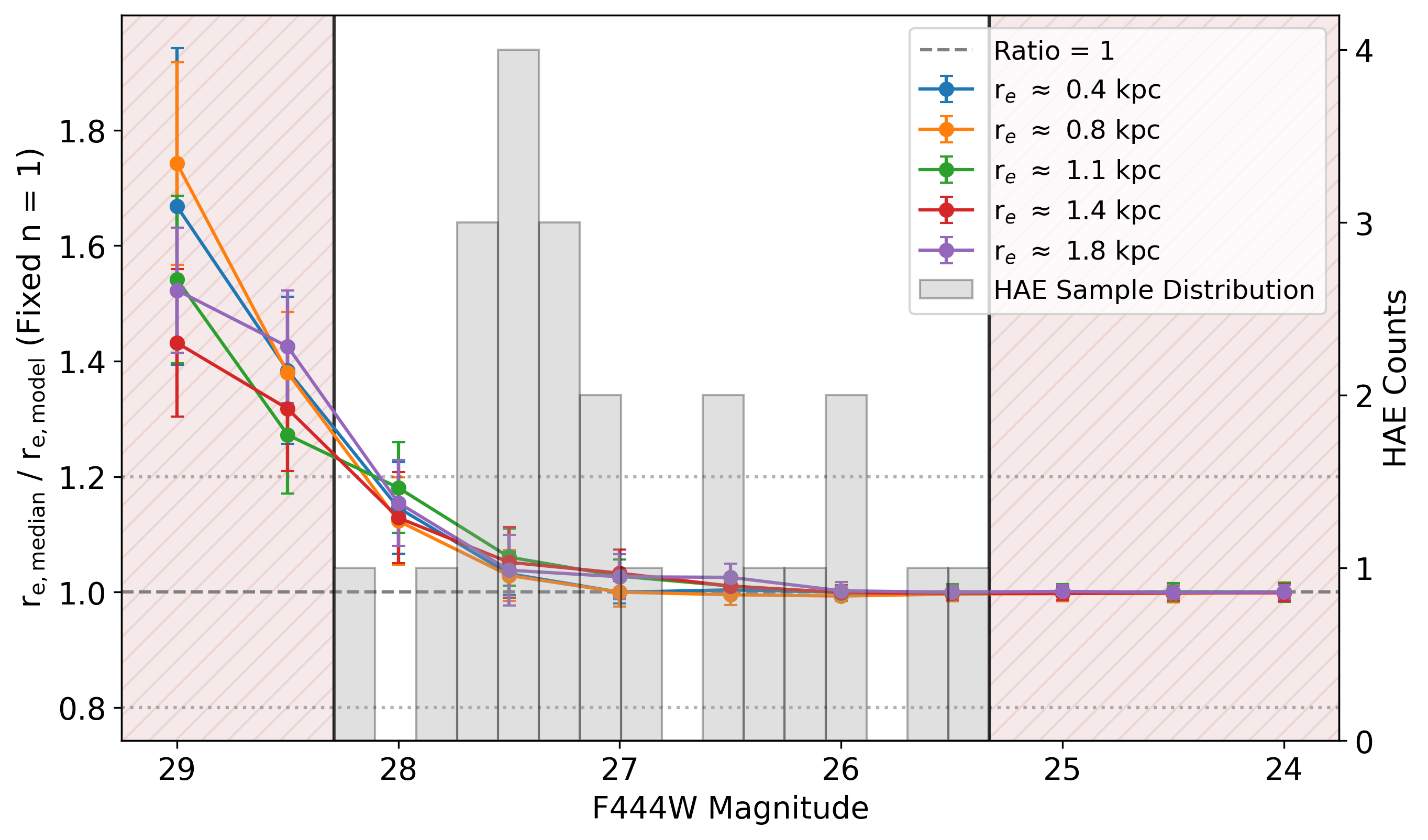}
    \caption[]{The ratio of median extracted $r_{e, \text{F444W}}$ to model values of mock galaxies (left axis) as a function of model F444W magnitude. For each magnitude, recovered $r_{e, \text{median}}$ are determined for mock galaxies with model radii of $\approx 0.4$\,\si{\kilo\parsec} (blue), $\approx 0.8$\,\si{\kilo\parsec} (orange), $\approx 1.1$\,\si{\kilo\parsec} (green), $\approx 1.4$\,\si{\kilo\parsec} (red) and $\approx 1.8$\,\si{\kilo\parsec} (purple) at $z=6.1$. The grey dashed line represents $r_{e, \text{median}}/r_{e, \text{expected}} = 1$, with the grey dotted lines representing $\pm\,0.2$. We represent the F444W magnitude range our actual sample of \ac{HAEs} with a grey histogram (right axis). The vertical black lines are the upper and lower bounds of the F444W magnitude range of our \ac{HAEs}, with the shaded brown regions indicating regions outside of that range.}
        \label{fig::galfiterrors}
\end{figure}

In Section \ref{subsec::galfitfitting}, we discussed the uncertainty estimations of \texttt{GALFIT} and how it typically underestimates them (see \citealp{Haussler2007,Haussler2013}; also \citealp{vanderWel2012}). There is also evidence that for faint, compact objects, \texttt{GALFIT} begins to overestimate the sizes of galaxies. For example, \citet{Davari2014} found that \texttt{GALFIT} can overestimate the sizes by as much as $20\%$ when fitting a single S\'ersic profile to multi-component, early-type galaxies (see also \citealp{Mosleh2013,Meert2013,Wang2024b}), though they note galaxies at high-$z$ like those we study in this paper are not as prone to those specific issues. Moreover, our \ac{HAEs} are likely late-type galaxies given their selection criteria and \ac{SFR} \citep{Pirie2025a}.

Despite this, we decided to probe \texttt{GALFIT}'s ability to recover accurate $r_{e}$ measurements by measuring the S\'ersic profiles of model galaxies in 33 000 mock images in F444W with similar properties to those of our sample of \ac{HAEs}. In total, we created 330 $n = 1$ model galaxies using \texttt{GALFIT} with properties in the range $m_{\text{F444W}} = 24 - 29$\,mag, $r_{e, \text{model}} \approx 0.4 - 1.8$\,\si{\kilo\parsec} and $b/a = 0.2 - 1.0$. Each mock galaxy was then placed in 100 random sky cutouts of the full PRIMER F444W image. These sky cutouts were created by ensuring that no {\sc SExtractor}-detected sources were within $\sqrt{100^{2} + 100^{2}} \approx 141$ pixels of the centre of the cutout. We then ran \texttt{GALFIT} on these model galaxies with sky backgrounds following the steps in Section \ref{subsec::galfitfitting} with $n$ kept fixed at $n = 1$, as well as fits where it is left as a free parameter for a total of 66 000 fits. We then bin the results on the model F444W magnitude and $r_{e, \text{model}}$, with each bin containing 600 outputs. We then remove any catastrophic fitting errors which we define here as those that run up against the $r_{e}$ constraint boundaries (or $n$ boundaries for when $n$ is a free parameter). In other words, we remove any fit that produces a fit with $r_{e} = 0.1$\,pixels or $r_{e} = 100$\,pixels ($n = 0.2$ or $n = 8$ for the free S\'ersic index fits). The number of catastrophic errors varied depending on model values, reaching as high as $67\%$ (402/600) for models with an apparent F444W magnitude of 29 mag and $r_{e} \approx 1.4$\,\si{\kilo\parsec}. From the remaining fits, we then determine the median recovered $r_{e}$ ($r_{e, \text{median}}$) in each bin and the standard error.

Figure \ref{fig::galfiterrors} shows the $r_{e, \text{median}}/r_{e, \text{model}}$ ratio (left axis) of fixed $n = 1$ mock galaxies in 0.5-wide magnitude bins. The different coloured points represent different $r_{e, \text{model}}$ values ranging from $\approx0.4$\,\si{\kilo\parsec} to $\approx1.8$\,\si{\kilo\parsec}. Overlaid is a histogram of the F444W magnitude counts (right axis) of our sample of 23 \ac{HAEs}, with vertical black lines indicating the upper and lower bounds of our sample. The shaded regions highlight magnitudes outside the range of our sample. From Figure \ref{fig::galfiterrors}, we see that for $n = 1$ galaxies at F444W magnitudes $\lesssim26.5$\,mag, the recovered $r_{e, \text{median}}$ are consistent with $r_{e, \text{model}}$ within uncertainties regardless of $r_{e, \text{model}}$. However, at fainter magnitudes, particularly at $\gtrsim28$\,mag, \texttt{GALFIT} consistently overestimates the sizes, reaching as high as $\sim74\%$ larger for $r_{e, \text{model}} \approx 0.8$\,\si{\kilo\parsec} at $29$ mag. This extreme is beyond the range of our sample, however, and all of the ratios for any $r_{e, \text{model}}$ within our magnitude range are $\lesssim20\%$ overestimation. This further justifies our floor uncertainty value of $25\%$ for our \ac{HAEs} in Section \ref{subsec::galfitfitting} as a conservative estimate. The general trend from Figure \ref{fig::galfiterrors} is that the fainter the magnitude, the more \texttt{GALFIT} overestimates the sizes of known models. We believe this is a result of the sky background becoming more indistinguishable from the faint edges of these objects causing \texttt{GALFIT} to calculate the $r_{e}$ of an object that extends further into the sky background than it does in the injected model. This overestimation may introduce some scatter of galaxy sizes at faint magnitudes (see Section \ref{subsec::scatter_discussion}).


\section{Results}
\label{sec::results}

\begin{figure*}
\centering
\begin{subfigure}{0.45\linewidth}
\includegraphics[width=\linewidth, trim=30 0 30 0]{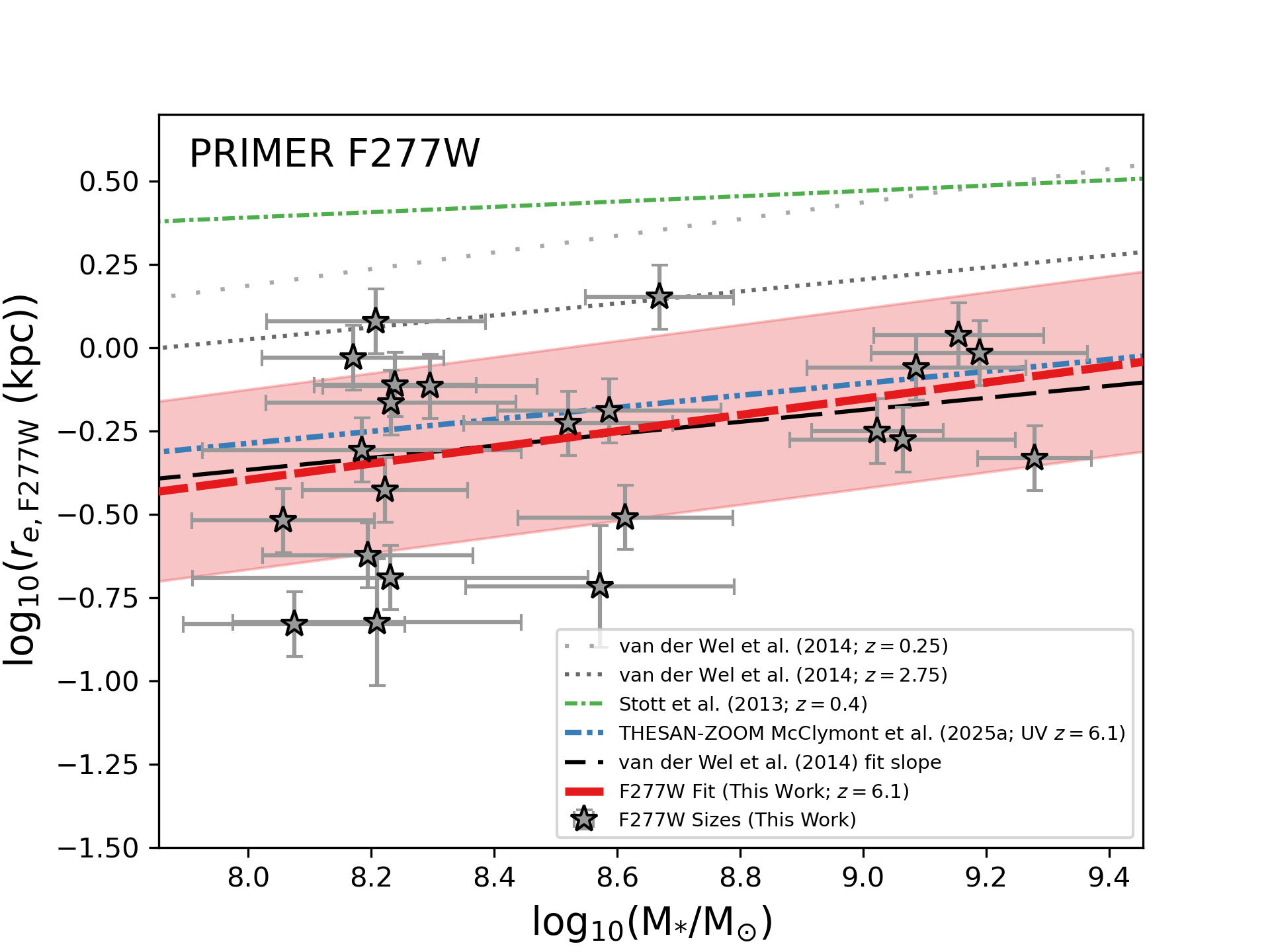}
\caption{ }
\label{subfig::uvsizemass}
\end{subfigure}
\hfill
\begin{subfigure}{0.45\linewidth}
\includegraphics[width=\linewidth, trim=30 0 30 0]{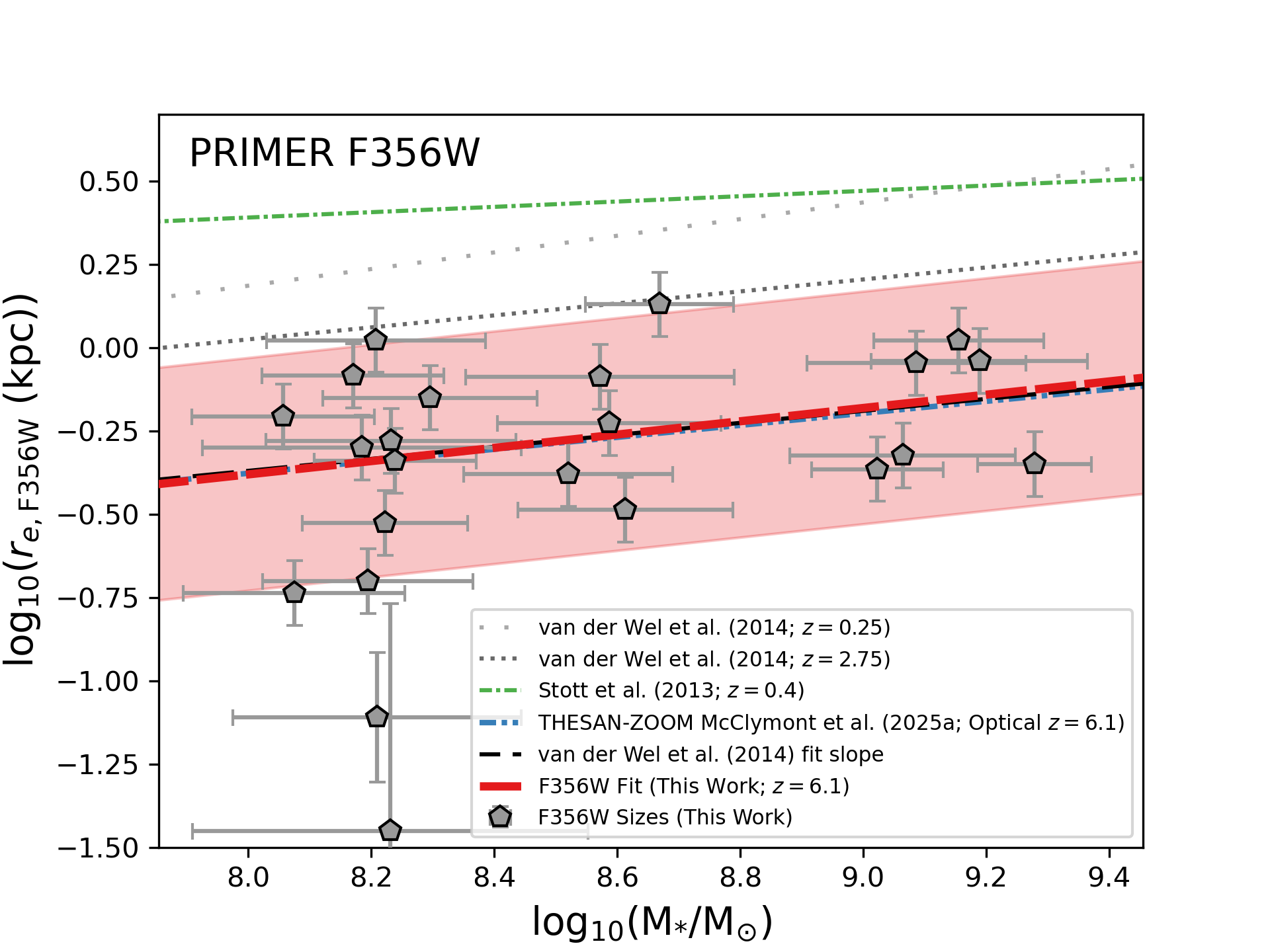}
\caption{ }
\label{subfig::vbandsizemass}
\end{subfigure}

\vspace{-0.1em} 

\begin{subfigure}{0.45\linewidth}
\includegraphics[width=\linewidth, trim=30 0 30 0]{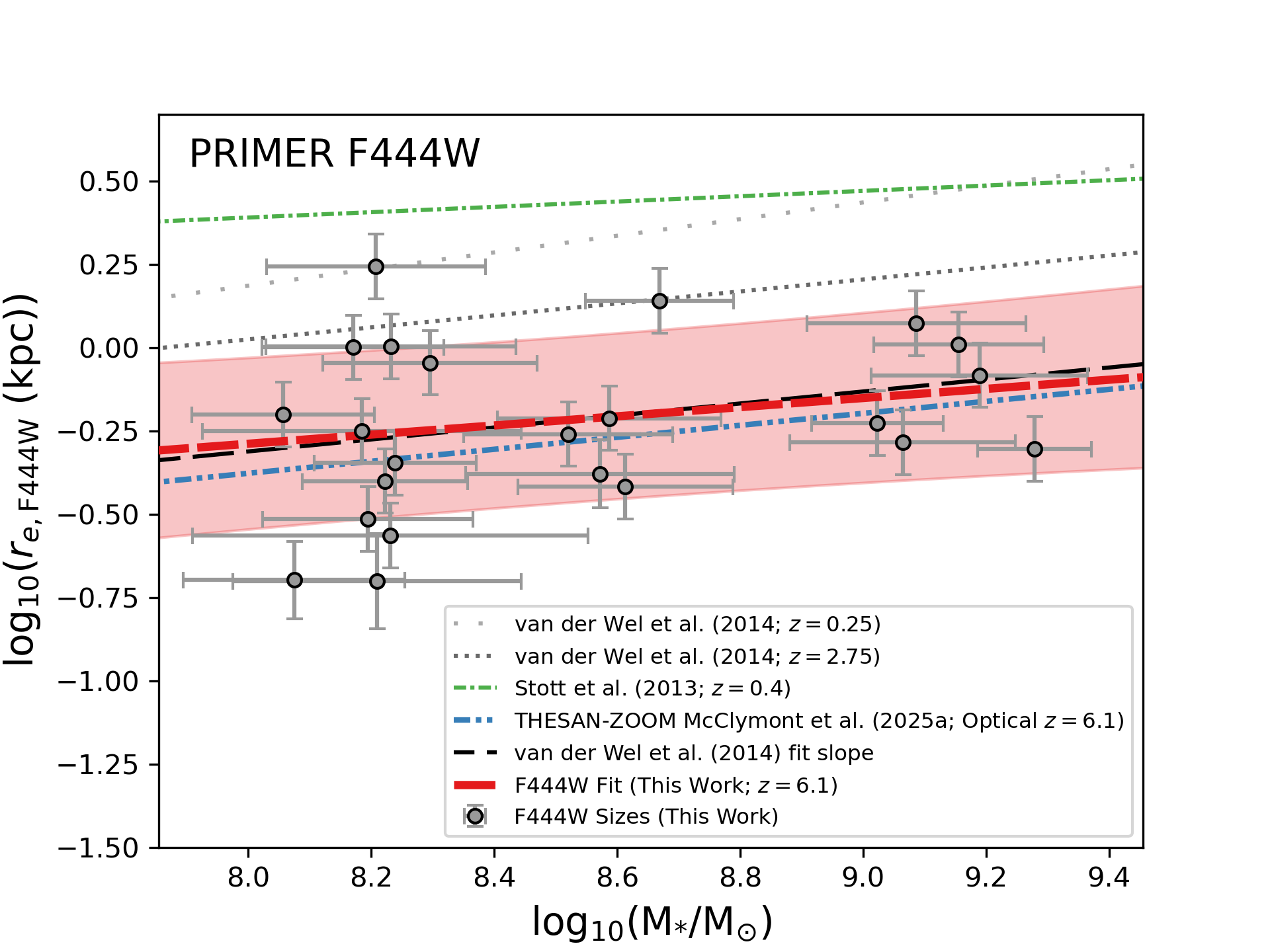}
\caption{ }
\label{subfig::bbsizemass}
\end{subfigure}
\hfill
\begin{subfigure}{0.45\linewidth}
\includegraphics[width=\linewidth, trim=30 0 30 0]{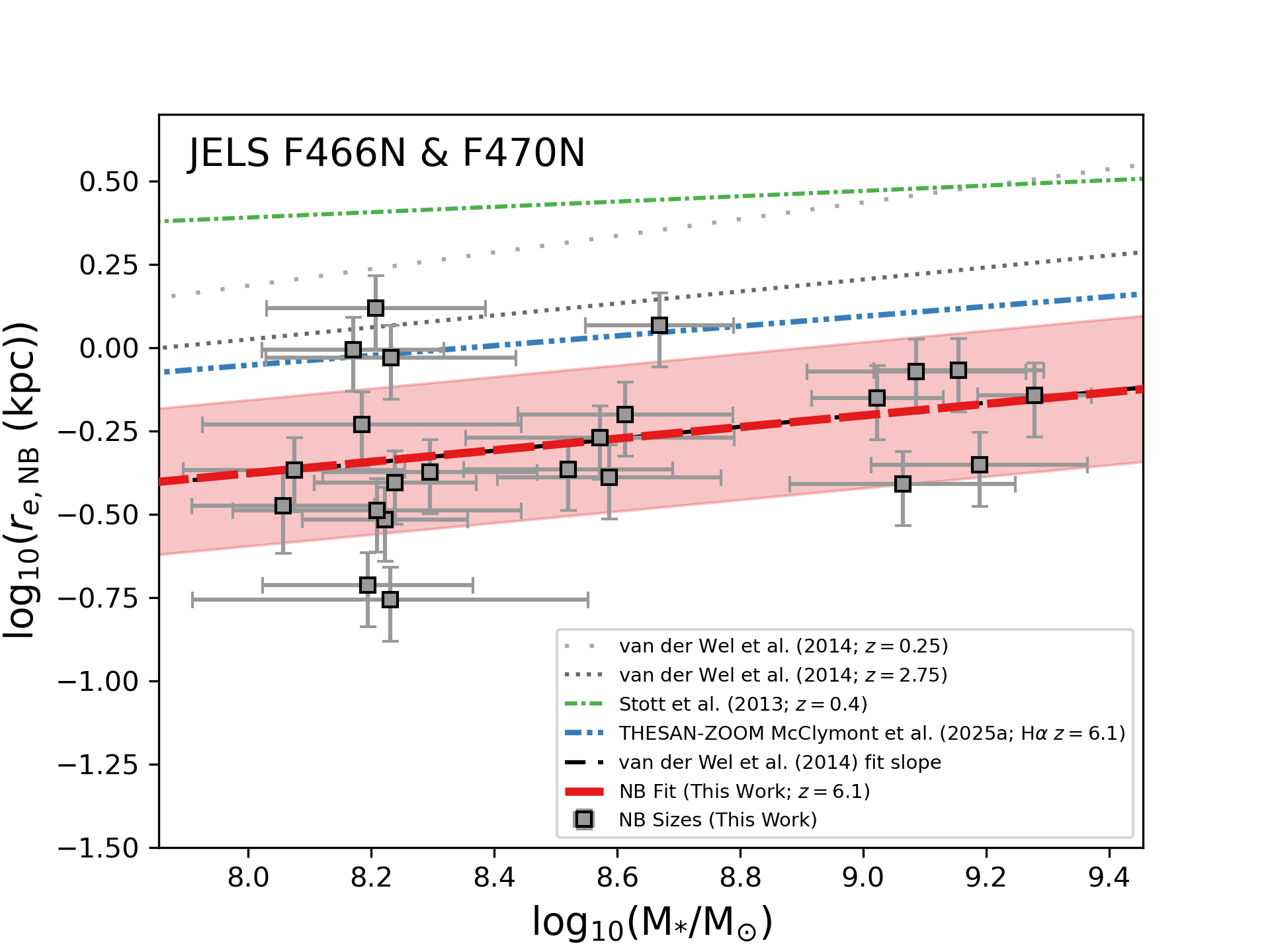}
\caption{ }
\label{subfig::nbsizemass}
\end{subfigure}

\caption{The $r_{e}-M_{*}$ relationship for our \ac{HAEs} at $z=6.1$ in four different \emph{JWST} bands: PRIMER F277W (rest-\ac{NUV}; \emph{Upper Left}), PRIMER F356W (rest-$V$-band; \emph{Upper Right}) and PRIMER F444W (rest-$R$-band; \emph{Lower Left}), and JELS F466N/F470N NB (rest-H$\alpha$; \emph{Lower Right}). The grey symbols represent the individual sizes of each HAE. In all panels, the red dashed line shows the fitted relationship to the individual points, with the red shaded region indicating the 1$\sigma$ scatter. We compare to the observed $r_{e}-M_{*}$ relationship of \ac{HAEs} from \citet{Stott2013a} at $z=0.4$ (green dash-dotted line), as well as those from \citet{vanderWel2014a} at $z=0.25$ (loosely dotted grey line) and $z=2.75$ (densely dotted grey line). We also compare to the simulated $r_{e}-M_{*}$ relationship from \citet{McClymont2025a} (blue dash-dotted) who employ the {\sc thesan-zoom} radiation-hydrodynamics zoom-in simulations \citep{Kannan2025}. \citet{McClymont2025a} measure the $r_{e}-M_{*}$ in three different bands: \ac{UV} (rest-frame $0.1475 - 1525$\,\si{\micro\metre}), optical (rest-frame $0.5 - 0.6$\,\si{\micro\metre}) and H$\alpha$. We select the appropriate relationship for comparison in each of our observed bands. The black long-dashed line shows the fixed slope of the \citet{vanderWel2014a} $r_{e}-M_{*}$ relationship at $z=2.75$ with the offset fitted to our data points.}
\label{fig::sizemassrelationship}
\end{figure*}

In this Section, we detail the key results from our analysis. We do this first by determining whether we observe a $r_{e}-M_{*}$ relationship at $z = 6.1$ and then comparing NB and F444W sizes. The latter allows us to compare the size of the H$\alpha$-selected \ac{SF} component of our \ac{HAEs} to the spatial extent of the established stellar component, inferred from the F444W photometry. Additionally, since both NB filters overlap with F444W, we also fit light profiles to our \ac{HAEs} with the modelled NB emission removed from the F444W image in order to account for H$\alpha$ contributions to the BB (see Section \ref{subsubsec::nb_in_bb_emission}). In Section \ref{subsec::redshiftevolution}, we will compare our measured $r_{e, \text{F444W}}$ to studies at a range of redshifts.


\subsection{Size-Mass Relationship}
\label{subsec::sizemasssec}

Figure \ref{fig::sizemassrelationship} shows the $r_{e}-M_{*}$ relationship for our \ac{HAEs} in the PRIMER F277W (Figure \ref{subfig::uvsizemass}), PRIMER F356W (Figure \ref{subfig::vbandsizemass}), PRIMER F444W (Figure \ref{subfig::bbsizemass}) and JELS F466N/F470N NB (Figure \ref{subfig::nbsizemass}) images. In each panel, the red dashed line indicates the best fit power-law of the form $\log_{10}(r_{e}\text{ / kpc}) = \alpha\log_{10}(M_{*}/\si{\solarmass}) + A$ to the individual $r_{e}$ points, determined using the \texttt{curve\_fit} function from the \texttt{scipy.optimize} module in Python \citep{Virtanen2020}. The shaded region indicates the 1$\sigma$ scatter at fixed stellar mass. The parameters for these fits can be found in Table \ref{tab::sizemassdata}. We compare to the $z=0.4$ $r_{e}-M_{*}$ relationship of \citet{Stott2013a} who analysed the structural properties of a sample of \ac{HAEs}, though we note that their $r_{e}$ measurements are determined from ground-based observations \citep{Geach2008,Sobral2013a}. We also compare to \citet{vanderWel2014a} for \ac{SF} galaxies at $z\sim0.25$ (rest-frame $Y$-band; light-grey dotted line) and $z\sim2.75$ (rest-frame $B$-band; dark-grey dotted line). Despite being at different rest-frame wavelengths, the observed $r_{e}-M_{*}$ relationships we compare to in Figure \ref{fig::sizemassrelationship} are also measured at wavelengths redward of the rest-4000\,\si{\angstrom} break so are less affected by ongoing star formation, thus making them reasonable comparisons. We note that these comparisons only apply to BB sizes but we include them on the NB plot for reference. Additionally, we look at how our $r_{e}-M_{*}$ relationship compares to those in simulations from \citet{McClymont2025a}. They measure the 2D half-light radii of galaxies in the {\sc thesan-zoom} radiation-hydrodynamics zoom-in simulations \citep{Kannan2025}, a high-resolution successor to the large-volume {\sc thesan} simulations \citep{Kannan2022}. After accounting biases to better match observations, \citet{McClymont2025a} measure the $r_{e}-M_{*}$ relationship in three different bands: \ac{UV} (rest-frame $0.1475 - 1525$\,\si{\micro\metre}), optical (rest-frame $0.5 - 0.6$\,\si{\micro\metre}) and H$\alpha$ emission. In Figure \ref{fig::sizemassrelationship}, we compare each relation from \citet{McClymont2025a} to the appropriate observed filter for our sizes. We will focus on the $r_{e}-M_{*}$ relationship in F444W (rest-frame $R$-band) for the rest of this study as it overlaps with the JELS F466N/F470N NB observations and gives a better reflection of the underlying stellar population than bluer BB or the NB data.

From Figure \ref{fig::sizemassrelationship}, we observe a $r_{e}-M_{*}$ relationship for \ac{HAEs} at $z=6.1$ with a slope of $\alpha_{\text{F444W}} = 0.14\pm0.12$ in F444W and $\alpha_{\text{H}\alpha} = 0.17\pm0.12$ in the NB data. The F444W $r_{e}-M_{*}$ relationship is significantly offset from those in both \citet{Stott2013a} and \citet{vanderWel2014a}, reflecting the accepted trend in the literature that, for fixed stellar mass, galaxies at higher redshifts have smaller $r_{e}$ (e.g. \citealp{Shibuya2015,Mowla2019b,Sun2024,vanderWel2024}; see Section \ref{subsec::redshiftevolution}). For a fixed stellar mass of $10^{9.25}$\,\si{\solarmass}, we find an offset in $\log_{10}(r_{e}\text{ / kpc})$ from the \citet{vanderWel2014a} $z = 2.75$ relationship of $-0.37\pm0.10$ ($-0.41\pm0.10$) dex for our F444W (NB) derived relationship. We choose to use a fixed stellar mass of $10^{9.25}$\,\si{\solarmass} despite being near the upper end of our sample because comparisons in the literature are difficult at lower stellar masses (see Section \ref{subsec::redshiftevolution}). The offset from \citet{vanderWel2014a} reflects an increase in average $r_{e}$ of $\approx1$\,\si{\kilo\parsec} from $z=6.1$ to $z=2.75$, or a factor of $\sim2.3-2.5$ increase in just $\sim1.4$\,\si{\giga\year}. According to \citet{vanderWel2014a}, from $z=2.75$ to $z=0.25$, the $r_{e}$ of a $10^{9.25}$\,\si{\solarmass} \ac{SF} galaxy increases by a factor of $\approx1.8$ in $\sim8.2$\,\si{\giga\year}, suggesting significantly more rapid galaxy growth before Cosmic Noon than after. This is also indicated by the near-identical value of the $z=0.4$ relationship found by \citet{Stott2013a} at this stellar mass. Similar offsets and slopes to these observational relationships are seen in all filters in Figure \ref{fig::sizemassrelationship}. Indeed, from Table \ref{tab::sizemassdata}, all of our $r_{e}-M_{*}$ relationships are consistent within errors, with weak evidence the slope may get shallower with increasing rest-frame wavelength, a trend that has been seen in the literature \citep{Nedkova2024,Allen2025,Jia2024,Yang2025}.

We find good agreement between our BB $r_{e}-M_{*}$ relations and the simulated $z=6.1$ results of \citet{McClymont2025a}, with all trends occupying the 1$\sigma$ scatter about the relationships. This agreement is particularly strong for our PRIMER F356W measurements, where the mean offset is only $\approx-0.01$ dex in half-light radius across our stellar mass range. In contrast, our $z=6.1$ H$\alpha$ $r_{e}-M_{*}$ relation disagrees with the \citet{McClymont2025a} prediction, with their $r_{e}$ values $\approx0.3$ dex larger. They interpret their large H$\alpha$ sizes as being due to nebula emission beyond the stellar and UV continuum as extreme Lyman-continuum emission from a central starburst region ionises gas reservoirs surrounding the galaxy. However, our observations do not support this scenario.

The slope of our F444W $r_{e}-M_{*}$ relationship ($\alpha_{\text{F444W}} = 0.14\pm0.12$) is consistent with those for late-type galaxies in \citet{vanderWel2014a} who find $\alpha=0.18\pm0.02$ at $z=2.75$ and $\alpha=0.25\pm0.02$ at $z=0.25$. We illustrate the consistency with \citet{vanderWel2014a} by fitting a line with a fixed slope equal to their $z=2.75$ relationship to our sample (black dashed line) and finding that it is within the $1\sigma$ scatter of our fit for our full stellar mass range. The large errors on our relationship are likely explained by the much-reduced sample size compared to \citeauthor{vanderWel2014a} (\citeyear{vanderWel2014a}; 23 vs $\sim2000$) and the large scatter of $\sigma_{\text{scatter}} = 0.30$ dex of our individual sizes at low stellar mass ($M_{*}<10^{8.4}$\,\si{\solarmass}; compared to $\sigma_{\text{scatter}} = 0.16$ dex at $M_{*} \geq 10^{8.4}$\,\si{\solarmass}). We discuss the possible causes of this increased scatter in Section \ref{subsec::scatter_discussion}. We also find that the slope of our $r_{e}-M_{*}$ relationship is consistent within errors with HAE the relationship of \citet{Stott2013a} at $z=0.4$ of $\alpha=0.03\pm0.02$. These consistencies, although caveated by large relative errors, suggest the trend in the literature that the late-type $r_{e}-M_{*}$ slope remains generally unchanged with redshift may continue out to $z=6.1$ (see also \citealp{Shen2023,Ito2023,Ward2024,Allen2025}), though there is evidence of steeper slopes at $z\lesssim0.1$ (e.g. \citealp{Shen2003,Guo2009,Paulino-Afonso2017}). When combined with our H$\alpha$ to stellar continuum size ratios (see Section \ref{subsec::bbtonbsizesec}), we believe that the lack of significant evolution in the $r_{e}-M_{*}$ slope is a consequence of \ac{SF} galaxies primarily building their mass through secular star formation across cosmic time. We explore this further in Section \ref{subsec::discussbbtonb}.

\begin{table}
    
    \footnotesize
    \centering
    \caption{$r_{e}-M_{*}$ relationships as seen in Figure \ref{fig::sizemassrelationship}. These fits are of the form $\log_{10}(r_{e}\text{ / kpc}) = \alpha\log_{10}(M_{*}/\si{\solarmass}) + A$. }
    \label{tab::sizemassdata}
    \begin{tabular}{lcc} 
        \hline
        Image & $\alpha$              & $A$           \\
        \hline
F277W                       & $0.24\pm0.13$  & $-2.34\pm1.11$ \\
F356W                       & $0.20\pm0.12$  & $-1.98\pm1.07$ \\
F444W                       & $0.14\pm0.12$  & $-1.39\pm1.06$ \\
NB                          & $0.17\pm0.12$  & $-1.77\pm1.01$ \\        \hline
F444W$_{\text{sub}}$        & $0.08\pm0.12$  & $-0.82\pm1.04$ \\        \hline{}
    \end{tabular}
\vspace{-4mm}
\end{table}


\subsection{Stellar Component to Star-Forming Region Size Ratio}
\label{subsec::bbtonbsizesec}

\begin{figure}
        \centering
        \includegraphics[width=\columnwidth, trim=0 0 0 0, clip=true]{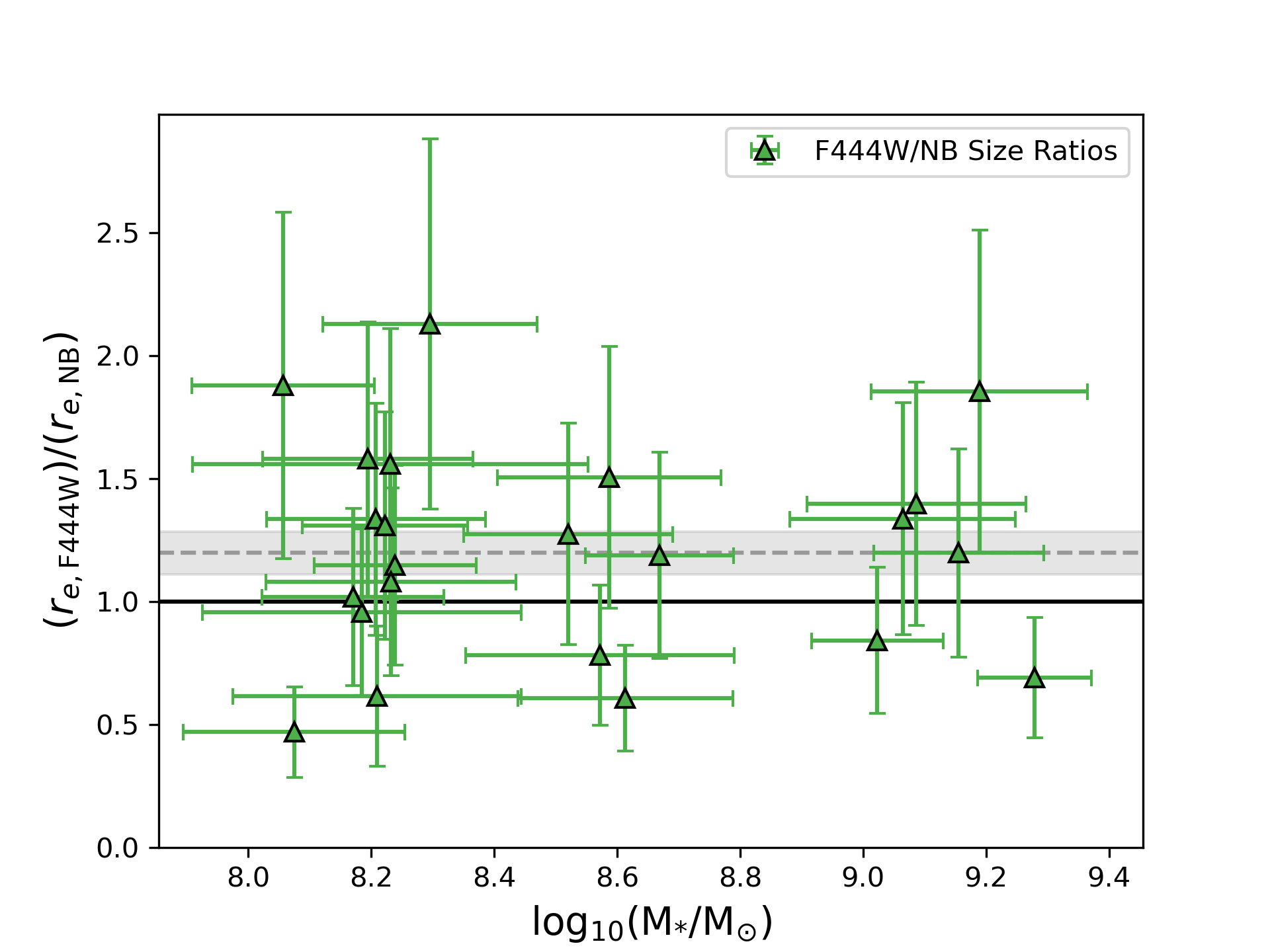}
    \caption[]{The ratio of the measured $r_{e}$ in F444W (rest-$R$-band; $r_{e, \text{F444W}}$) to the measured $r_{e}$ in NB ($r_{e, \text{NB}}$) for each of our \ac{HAEs} (green triangles) against stellar mass. The error on each size ratio represents the combined error on the respective $r_{e}$ measurements. The solid black line represents $r_{e, \text{F444W}}/r_{e, \text{NB}} = 1$. The dashed grey line represents the median $r_{e, \text{F444W}}/r_{e, \text{NB}} = 1.20$, with the shaded region indicating the standard error ($\pm\,0.09$).}
        \label{fig::bb_to_nb_size_ratio}
\end{figure}

Figure \ref{fig::bb_to_nb_size_ratio} shows the ratio of $r_{e, \text{F444W}}$ to $r_{e}$ measured in NB ($r_{e, \text{NB}}$) for each of our \ac{HAEs}. This ratio reflects the size of any stellar component compared to the \ac{SF} region traced by H$\alpha$ emission from H II regions surrounding young, massive stars. We find a median size ratio of $\frac{r_{e, \text{F444W}}}{r_{e, \text{NB}}} = 1.20\pm0.09$, indicating that the stellar emission is marginally larger than the H$\alpha$-emitting \ac{SF} component at the \ac{EoR}, suggestive of a more centrally concentrated \ac{SF} regions in \ac{HAEs} at $z = 6.1$. However, the uncertainty on the measurement of $\frac{r_{e, \text{F444W}}}{r_{e, \text{NB}}}$ for many of the individual galaxies is such that the ratio is consistent with 1. Therefore, we can say more broadly that the ratios in Figure \ref{fig::bb_to_nb_size_ratio} indicate that there are already-established stellar components in \ac{SF} galaxies at $z=6.1$ that are at least comparable to, if not larger than, the size of the expected \ac{SF} regions. This contrasts with the results of \citet{Nelson2016} who find that the active star formation traced by H$\alpha$ at $z = 0.7-1.5$ extends further than the existing stellar continuum. They conclude that their results show \ac{SF} galaxies at their redshift range are growing in size primarily from star formation (see also \citealp{Nelson2012,Matharu2022,Shen2024b}). \citet{Wilman2020} see a similar result, finding the median (mean) H$\alpha$ size being a factor of 1.18 (1.26) larger than the stellar continuum from their sample of $0.7<z<2.7$ observations in the KMOS$^{\text{3D}}$ survey \citep{Wisnioski2015,Wisnioski2019}.

\begin{figure}
        \centering
        \includegraphics[width=\columnwidth, trim=0 0 0 0, clip=true]{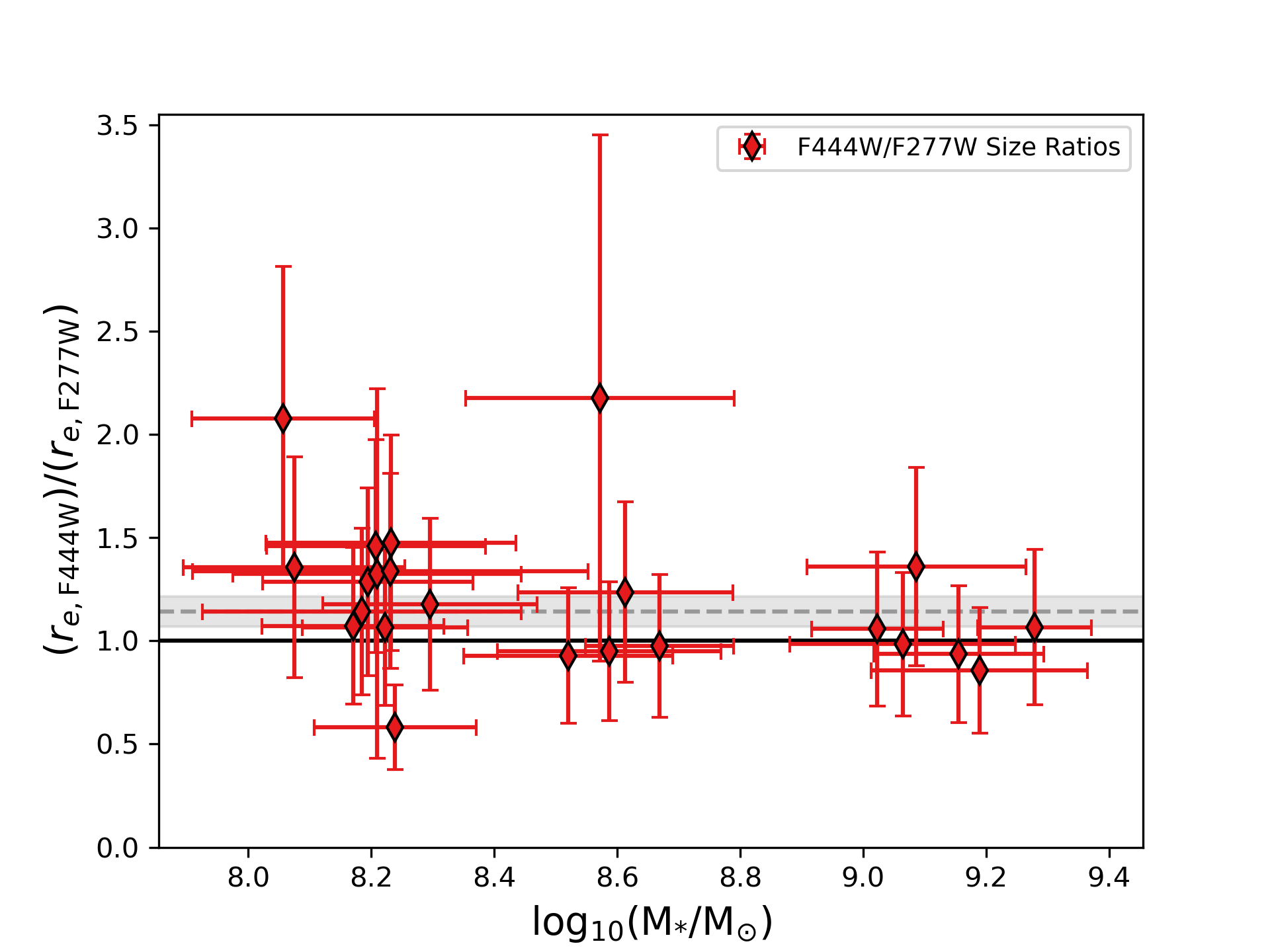}
    \caption[]{As in Figure \ref{fig::bb_to_nb_size_ratio}, but for our $r_{e, \text{F444W}}$ to $r_{e}$ measured in F277W (rest-\ac{NUV}; $r_{e, \text{F277W}}$) ratios (thin red diamonds). The dashed grey line and shaded region represent the median $r_{e, \text{F444W}}/r_{e, \text{F277W}} = 1.14 \pm 0.07$.}
        \label{fig::bb_to_uv_size_ratio}
\end{figure}

As briefly discussed in Section \ref{sec::intro}, the \ac{UV}- or \ac{NUV}-continuum are other frequently used indicators of star formation. Therefore, another method of measuring the extent of any established stellar component to \ac{SF} regions is to measure the ratio of $r_{e, \text{F444W}}$ to $r_{e}$ measured in F277W (rest-\ac{NUV}; $r_{e, \text{F277W}}$). We show this in Figure \ref{fig::bb_to_uv_size_ratio} where we find a median $\frac{r_{e, \text{F444W}}}{r_{e, \text{F277W}}}$ ratio of $1.14\pm0.07$. This agrees with our median $\frac{r_{e, \text{F444W}}}{r_{e, \text{NB}}}$ from Figure \ref{fig::bb_to_nb_size_ratio} and further suggests that the \ac{SF} region of our \ac{HAEs} is more centrally concentrated with an established stellar component that may extend beyond this. This reduced value could also be partly caused by \ac{UV} light being more affected by dust than H$\alpha$ emission.

Our results imply that, prior to the current period of star formation we are seeing traced by the H$\alpha$ emission in the NB data, there must have been a significant-enough episode of star formation to form a stellar component with a larger associated $r_{e}$. We discuss the implications and the possible causes of this in Section \ref{sec::discussion}.


\subsubsection{H$\alpha$ Contribution to F444W}
\label{subsubsec::nb_in_bb_emission}

The nature of NB imaging selection for detecting \ac{HAEs} mean that there could be a significant contribution from the H$\alpha$ emission line in the overlapping BB emission (in our case contributions to F444W from F466N or F470N). The median observed H$\alpha$ \ac{EW} for our sample is \ac{EW}$_{\text{H}\alpha} = 748 \pm 89$\,\si{\angstrom} so we decided to run analysis of our HAE sizes where the H$\alpha$ emission is removed from the F444W image to leave a NB flux-subtracted F444W (F444W$_{\text{sub}}$) image. To do this, we used the \texttt{GALFIT} output models from the NB fitting and subtracted them from the corresponding F444W image cutouts. This subtraction was done by scaling the NB flux density based on the relative effective widths of the F444W, F466N and F470N filters. The full subtraction is described by

\begin{equation}
    f_{\lambda, \text{F444W}_{\text{sub}}} = \frac{f_{\lambda, \text{F444W}} - f_{\lambda, \text{NB}}\left(\frac{W_{\text{eff, NB}}}{W_{\text{eff, F444W}}}\right)}   {1 - \frac{W_{\text{eff, NB}}}{W_{\text{eff, F444W}}}},
\label{eq::nb_subtracted_bb}{}
\end{equation}

\vspace{5pt}
\noindent where $f_{\lambda}$ and $W_{\text{eff}}$ are the flux density and effective width of a given filter respectively \citep{Waller1990}. In our case, $f_{\lambda, \text{F444W}}$ is the flux density of the F444W cutout of our \ac{HAEs}, and $f_{\lambda, \text{NB}}$ is the flux density of the \texttt{GALFIT} model output in NB, where NB is either F466N or F470N depending on which image the HAE was detected in (see middle panels of Figure \ref{fig::examplemodels} for example outputs). This method of model subtraction ensures that we are only removing the H$\alpha$ emission from the source without increasing the noise in the sky background, which \texttt{GALFIT} needs for accurate light profiles \citep{Peng2010a}. Once the NB models have been subtracted from the F444W cutouts, we ran \texttt{GALFIT} on the resulting images following the same steps as Section \ref{subsec::galfitfitting}.

\begin{figure}
        \centering
        \includegraphics[width=\columnwidth, trim=0 0 0 0, clip=true]{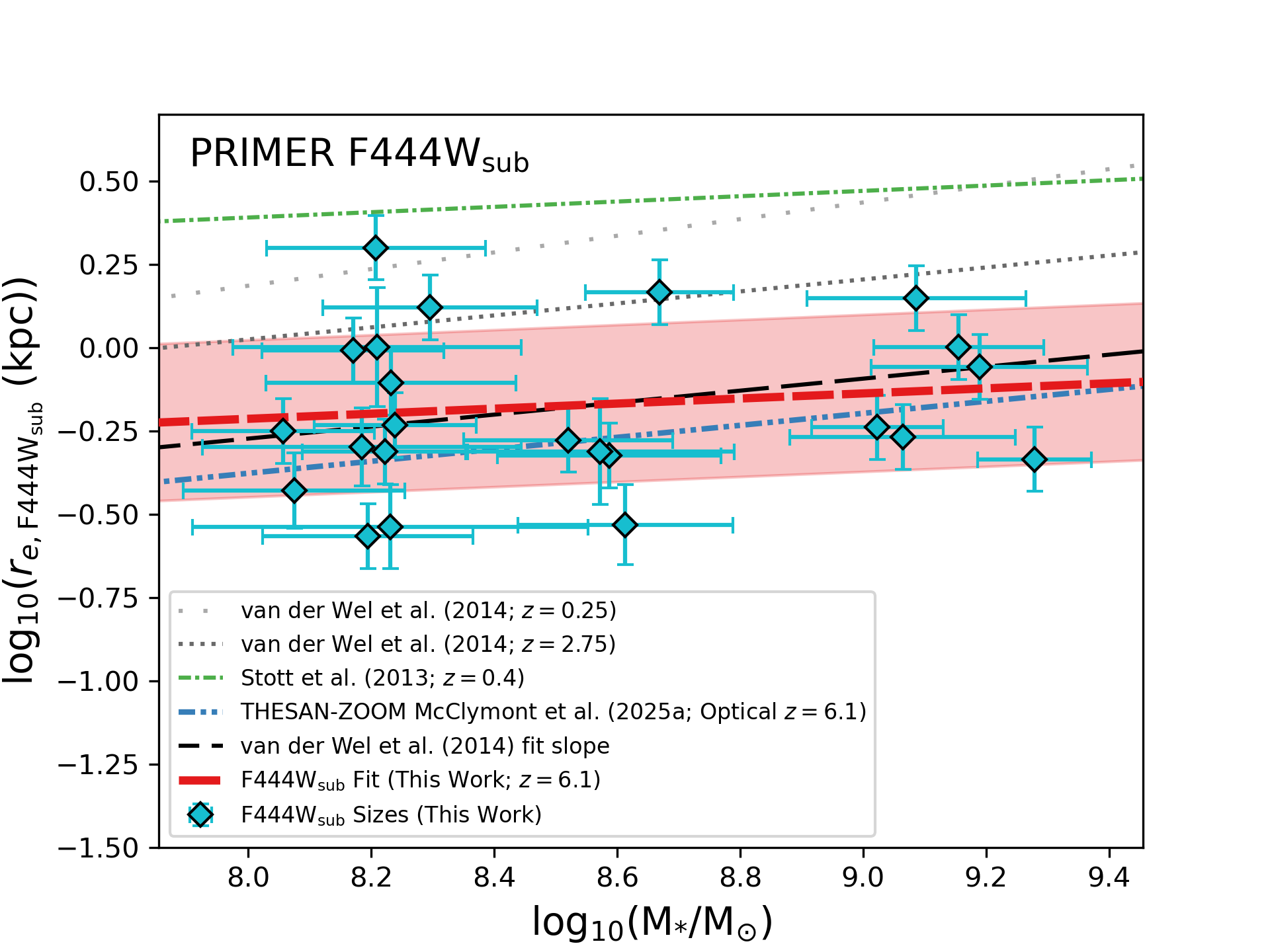}
    \caption[]{The $r_{e}-M_{*}$ relationship as in Figure \ref{fig::sizemassrelationship} for NB flux-subtracted F444W sizes ($r_{e, \text{F444W}_{\text{sub}}}$; cyan diamonds). The slope here is measured as $\alpha_{\text{F444W}_{\text{sub}}} = 0.08\pm0.12$ which is shallower than the slope seen measured in Figure \ref{subfig::bbsizemass} driven by an increase in $r_{e, \text{F444W}_{\text{sub}}}$ at $M_{*} < 10^{8.4}$\,\si{\solarmass}.}
        \label{fig::nb_subtracted_size_mass}
\end{figure}

\begin{figure}
        \centering
        \includegraphics[width=\columnwidth, trim=0 0 0 0, clip=true]{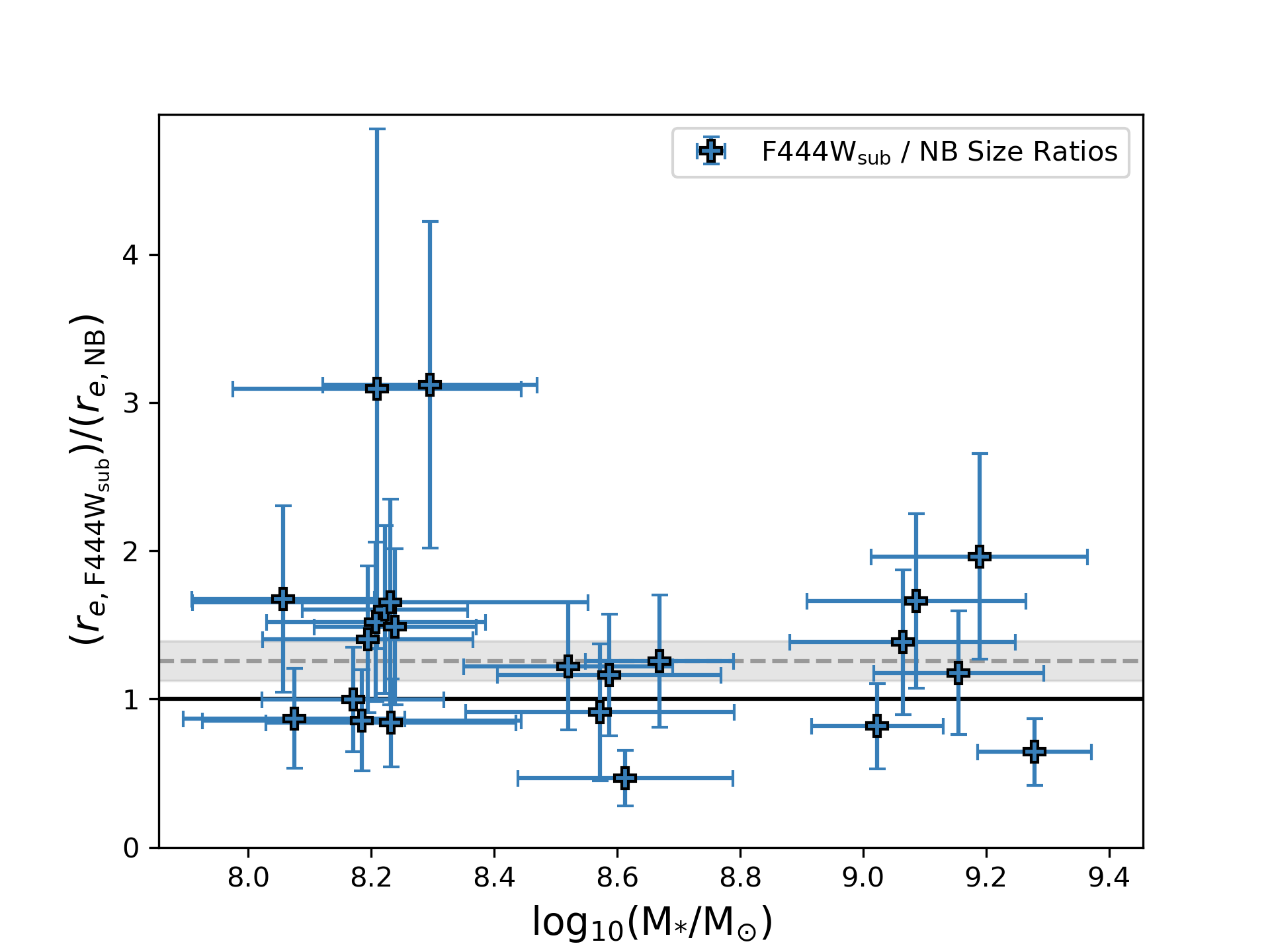}
    \caption[]{As in Figure \ref{fig::bb_to_nb_size_ratio}, but for our NB flux-subtracted F444W $r_{e}$ ($r_{e, \text{F444W}_{\text{sub}}}$) to $r_{e, \text{NB}}$ ratios (blue pluses). The dashed grey line and shaded region represent the median $r_{e, \text{F444W}_{\text{sub}}}/r_{e, \text{NB}} = 1.26 \pm 0.14$, which is marginally larger than than the $r_{e, \text{F444W}}/r_{e, \text{NB}} = 1.20 \pm 0.09$ seen in Figure \ref{fig::bb_to_nb_size_ratio}, though consistent within error.}
        \label{fig::nb_subtracted_size_ratios}
\end{figure}

Figure \ref{fig::nb_subtracted_size_mass} shows the $r_{e}-M_{*}$ relationship for our \ac{HAEs} in the same format as Figure \ref{fig::sizemassrelationship}, but using NB flux-subtracted $r_{e}$ ($r_{e, \text{F444W}_{\text{sub}}}$). We find that the slope of this relationship is shallower than those found in Figure \ref{fig::sizemassrelationship} (see Table \ref{tab::sizemassdata}) with $\alpha_{\text{F444W}_{\text{sub}}} = 0.08\pm0.12$. However, this is well within 1$\sigma$ of the previous relationships, as well as those from \citet{Stott2013a} and \citet{vanderWel2014a}.

We measure the ratio of $r_{e, \text{F444W}_{\text{sub}}}$ to $r_{e, \text{NB}}$ in Figure \ref{fig::nb_subtracted_size_ratios}. This gives us a cleaner comparison between the size of any stellar components and the \ac{SF} regions because we have removed contributions from the latter to the BB continuum. We find that the median size ratio when subtracting H$\alpha$ emission increases marginally to $\frac{r_{e, \text{F444W}_{\text{sub}}}}{r_{e, \text{NB}}} = 1.26 \pm 0.14$ (grey dashed line in Figure \ref{fig::nb_subtracted_size_ratios}), compared to $\frac{r_{e, \text{F444W}}}{r_{e, \text{NB}}} = 1.20\pm0.09$ without any subtraction, although the change is not significant. While this is a weak increase within the respective uncertainties, this marginal increase is in line with the \ac{SF} region traced by H$\alpha$ emission being more centrally concentrated.

This average increase in $r_{e}$ may contribute to the flattening of our $r_{e}-M_{*}$ slope in Figure \ref{fig::nb_subtracted_size_mass} compared to Figure \ref{subfig::bbsizemass}, particularly if the increase is predominately in low mass \ac{HAEs}. After subtracting H$\alpha$ emission, our $<10^{8.4}$\,\si{\solarmass} \ac{HAEs} have a ratio of $\frac{r_{e, \text{F444W}_{\text{sub}}}}{r_{e, \text{F444W}}} = 1.10 \pm 0.32$ compared to $\frac{r_{e, \text{F444W}_{\text{sub}}}}{r_{e, \text{F444W}}} = 0.98 \pm 0.04$ for $\geq10^{8.4}$\,\si{\solarmass}. The larger errors at $<10^{8.4}$\,\si{\solarmass} mean it is difficult to draw any definitive conclusions, and the ratio of both $r_{e, \text{F444W}}$ and $r_{e, \text{F444W}_{\text{sub}}}$ to $r_{e, \text{NB}}$ are consistent with each other, which suggests the overall sizes are not significantly affected by H$\alpha$ emission. This is not surprising, since our median observed \ac{EW}$_{\text{H}\alpha}\sim750$\,\si{\angstrom} is $\approx7\%$ the width of the F444W filter ($W_{\text{eff}} = 10676$\,\si{\angstrom}).

\subsection{Redshift Evolution of Galaxy Sizes}
\label{subsec::redshiftevolution}

\begin{figure*}
        \centering
        \includegraphics[width=\textwidth, trim=0 0 0 0, clip=true]{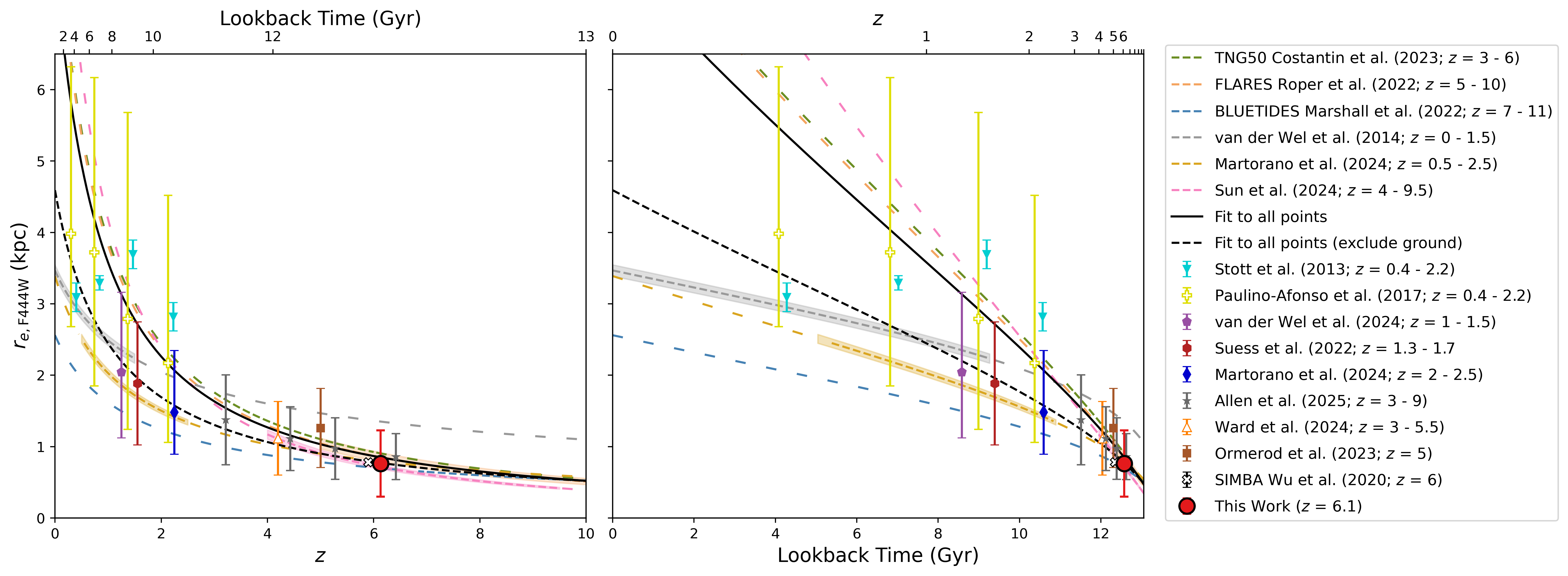}
    \caption{\emph{Left} - Size-$z$ relationship of studies in the literature compared to our work. \emph{Right} - The same as in the left panel, but all points and relations are as a function of lookback time. The large red circle indicates the size of a $10^{9.25}$\,\si{\solarmass} \ac{SF} galaxy for our sample derived from our F444W $r_{e}-M_{*}$ relationship in Figure \ref{subfig::bbsizemass}. The error associated with this size is the scatter about the relationship at this stellar mass. For the individual points, filled faces indicate that the inferred size is derived from a study with a stellar mass (or stellar mass range) that is within the range of this work. White faces indicate that the inferred mass has been extrapolated outside the mass range of that study to match the $10^{9.25}$\,\si{\solarmass} we use for our own estimate. We refer the reader to Table \ref{tab::sizezsymbols} for information on these studies. From observations, we compare to the size-$z$ relationships of \citet{vanderWel2014a} (grey; $0<z<1.5$), \citet{Martorano2024} (yellow; $0.5<z<2.5$) and \citet{Sun2024} (pink; $4<z<9.5$). From simulations, we compare to the size-$z$ relationship from the TNG50 cosmological hydrodynamical simulation \citep{Nelson2019} in \citet{Costantin2023} (green; $3<z<6$), the FLARES zoom-in simulations \citep{Vijayan2021} in \citet{Roper2022} (orange; $5<z<10$) and the BlueTides cosmological hydrodynamical simulations \citep{Feng2016} analysed in \citet{Marshall2022} (blue; $7<z<11$). The shaded regions of the literature relationships are the $1\sigma$ scatter in their relationships within their $z$ range (if applicable). Where the dashed lines become more spaced is an extrapolation beyond the redshift of the respective study. The solid black line is a fit to each of the individual points, weighted by their errors. We also include a fit that does not include the ground-based observations from \citet{Stott2013a} and \citet{Paulino-Afonso2017} (dashed black line). The points from \citet{Paulino-Afonso2017} and \citet{Wu2020a} have been shifted in both redshift (-0.1; \emph{left}) and lookback time (-0.2\,\si{\giga\year}; \emph{right}) for clarity.}
    \label{fig::redshiftandtime}
\end{figure*}

The narrow wavelength range probed by the F466N and F470N filters dictates that the redshift range we can probe for our \ac{HAEs} is similarly narrow ($6 \lesssim z \lesssim 6.2$, or $\approx0.04$\,\si{\giga\year} of cosmic time). As a result, we cannot model the redshift evolution of galaxy sizes across the \ac{EoR} ($6\lesssim z \lesssim 15$; \citealp{Fan2006b,Robertson2013}). Instead, we can see how the results from our unbiased, rest-optical \ac{HAEs} compare to observations in the literature \citep{Stott2013a,vanderWel2014a,Paulino-Afonso2017,vanderWel2024,Suess2022,Ormerod2023,Allen2025,Martorano2024,Sun2024,Ward2024}, as well as predictions from simulations \citep{Wu2020a,Roper2022,Marshall2022,Costantin2023}. These studies measure $r_{e}$ of \ac{SF} galaxies in different ways and we will briefly outline the data of each individual study, all of which are plotted in Figure \ref{fig::redshiftandtime}. We note here that all the observations in the referenced literature measure $r_{e}$ in rest-frame optical bands that are redward of the 4000\,\si{\angstrom} break.

As discussed in Section \ref{sec::intro}, \citet{vanderWel2014a} analysed the mass-size relation of galaxies between $0 \lesssim z \lesssim 3$ from 3D-HST and CANDELS. Here, we look at the size-$z$ relation they find for \ac{SF} galaxies from their results, as well as individual $r_{e}$ values for $10^{9.25}$\,\si{\solarmass} \ac{SF} galaxies at $z=0.25$ and $z=2.75$ derived from their $r_{e}-M_{*}$ relationships. Note that the derived $r_{e}$ values are extrapolations, as \citet{vanderWel2014a} only fit their late-type galaxy relation for $\gtrsim 10^{9.48}$\,\si{\solarmass}. We also compare to the \ac{SF} $r_{e}-M_{*}$ relation in \citet{vanderWel2024}, who combine observations from \emph{JWST} NIRCam in CEERS combined with CANDELS \emph{HST} imaging. We compare to their median $r_{e}$ for $10^{9.2}$\,\si{\solarmass} galaxies at $z=1.0-1.5$ in rest-frame $0.5$\,\si{\micro\metre}. \citet{Suess2022} used data from CEERS and 3D-HST to measure $r_{e}$ in the F444W and F150W BB \emph{JWST} NIRCam filters, with stellar masses measured by \citet{Skelton2014}. For our comparison, we use the rest-frame $R$-band median size of their $10^{9-9.5}$\,\si{\solarmass} \ac{SF} galaxies at $z=1.3-1.7$ (median stellar mass $\approx 10^{9.22}$\,\si{\solarmass}) . We define a \ac{SF} galaxy from their sample using a $U - V<1.0$ colour cut (to distinguish them from passive galaxies) for galaxies that satisfy the \citet{Skelton2014} \say{use} flag = 1, which they define as a galaxy with photometry of reasonably uniform quality. For comparison to \citet{Ormerod2023}, we used the median size of $z=5$ disk-like galaxies derived from their size-$z$ relationship. \citet{Ormerod2023} develop their relationship based on CEERS observations in an overlapping region in the CANDELS field, with the median $r_{e}$ at $z=5$ being measured in F356W of \emph{JWST} NIRCam (rest-frame $R$-band). \citet{Allen2025} measures galaxy sizes from public data from CEERS, PRIMER-UDS and PRIMER-COSMOS, accessible via the DAWN \emph{JWST} Archive (DJA\footnote{\url{https://dawn-cph.github.io/dja/index.html}}; see \citealp{Valentino2023}). We look at the median $r_{e}$ they measure in F444W at the four median redshifts they list in Table A.1 of their paper (rest-frame $0.59-1.04$\,\si{\micro\metre}). From \citet{Martorano2024}, we compare to both the size-$z$ relation and the derived $r_{e}$ of $\log_{10}({M_{*, \text{median}}}/\text{M}_{\odot}) \approx 9.27$ \ac{SF} galaxies from their $r_{e}-M_{*}$ relationship at $z=2 - 2.5$. They measured the rest-frame 1.5\,\si{\micro\metre} $r_{e}$ for galaxies in COSMOS-WEB \citep{Casey2023} and PRIMER-COSMOS. \citet{Sun2024} used data from CEERS to measure the $r_{e}$ of \ac{SF} galaxies, fitting two-dimensional parametric models in seven \emph{JWST} NIRCam filters in both short-wavelength (SW) and LW channels. We compare to the size-$z$ relationship that they derive from their combined rest-frame optical ($\approx0.41 - 0.66$\,\si{\micro\metre}) measurements at $z=4 - 9.5$. The final observational result we compare to comes from \citet{Ward2024} who also used imaging from CEERS and CANDELS to measure $r_{e}$ of galaxies at rest-frame 0.5\,\si{\micro\metre}. We extrapolated their $r_{e}-M_{*}$ relation to derive the $r_{e}$ of a typical $10^{9.25}$\,\si{\solarmass} \ac{SF} galaxy at $z=3 - 5.5$.

As with the observations above, we also looked at predictions from various simulations. Firstly, we compared to the rest-frame optical $r_{e}$ estimates from the SIMBA cosmological hydrodynamical simulations \citep{Dave2019} as reported by \citet{Wu2020a}. We derived the $r_{e}$ of a typical $10^{9.5}$\,\si{\solarmass} \ac{SF} galaxy from their size-luminosity relations in SIMBA-25 assuming a \citet{Calzetti2000} dust-law and their S\'ersic fit method (see their Section 3.4). We note that the sizes reported by \citet{Wu2020a} assume that dust tracks metals and does not assume radiative transfer. We compare to the size-$z$ relation from \citet{Roper2022} as measured in the First Light And Reionisation Epoch Simulations (FLARES; \citealp{Lovell2020,Vijayan2021}) - a suite of zoom simulations based on the cosmological hydrodynamical simulations from the Evolution and Assembly of GaLaxies and their Environments (EAGLE: \citealp{Crain2015}) project. \citet{Roper2022} constrain the size-$z$ relationship at $z=5-10$ using rest-\ac{UV} size measurements. Their size-$z$ relation is based on sizes derived from their non-parametric pixel-based method which they conclude is robust at high-$z$ ($z\gtrsim5$; see their Section 4.2.2). Despite being primarily based on \ac{UV} sizes, we note that we find the ratio of $r_{e, \text{F444W}}/r_{e, \text{\ac{UV}}} = 1.14 \pm 0.07$, suggesting that the $V$-band emission is slightly larger than the size inferred from the \ac{UV} continuum, so comparisons to \citet{Roper2022} should be noted with caution. We also compare to the size-$z$ relationship modelled by \citet{Costantin2023} at $z=3-6$ in the Illustris TNG50 cosmological hydrodynamical simulation \citep{Torrey2019,Nelson2019} based on rest-frame optical measurements at $\approx0.51-0.89$\,\si{\micro\metre}. Finally, we look at the BlueTides cosmological hydrodynamical simulations in \citet{Marshall2022}. The size-$z$ model we use from them is constrained at $z=7-11$ based on rest-frame \ac{FUV}, however they find that their \ac{FUV} and optical sizes are similar so we decided to keep the comparison.

\begin{table*}
    
    \footnotesize
    \centering
    \caption{The studies that make up the individual points in Figure \ref{fig::redshiftandtime}.}
    \label{tab::sizezsymbols}
    \begin{tabular}{c|c|c|c}
    \hline
    Reference & $z$ & Rest-frame Wavelength (\si{\micro\metre}) & Stellar Mass ($\log_{10}({M_{*}}/\text{M}_{\odot})$) \\ \hline
    \citet{Stott2013a}    & 0.4 - 2.23 & 0.47 - 1.57 & 9.25 \\
    \citet{Paulino-Afonso2017}     & 0.4 - 2.23 & 0.25 - 0.57 & 9.74 - 9.96 \\
    \citet{vanderWel2024}    & 0.5 - 2.3 & 0.46 - 0.57 & 9.2 \\
    \citet{Suess2022}    & 1.3 - 1.7 & 0.55 - 0.65 & 9.22 \\
    \citet{Martorano2024}     & 2.0 - 2.5 & 1.24 - 1.45 & 9.27 \\
    \citet{Allen2025}    & 3.0 - 9.0 & 0.39 - 0.62 & 9.25 \\
    \citet{Ward2024}    & 3.0 - 5.5 & 0.42 - 0.58 & 9.25 \\
    \citet{Ormerod2023}    & $5.0$ & 0.48 - 0.76 & 9.3 - 11.1 \\
    \citet{Wu2020a}    & $6.0$ & 0.62 & 9.25 \\ \hline
    \end{tabular}
\end{table*}

In Figure \ref{fig::redshiftandtime}, we show how the $r_{e, \text{F444W}}$ of a $10^{9.25}$\,\si{\solarmass} \ac{SF} galaxy from our $r_{e}-M_{*}$ relationship (Figure \ref{subfig::bbsizemass}) compares to the studies mentioned above. We chose to use a stellar mass of $10^{9.25}$\,\si{\solarmass} for these comparisons rather than our median stellar mass of $M_{*, \text{median}} = 10^{8.30}$\,\si{\solarmass} because the majority of the observational studies derive their size-$z$ evolution models based on much higher characteristic masses. This value is still within our sample’s mass range, and is above the mass range where we see the largest scatter in sizes (see Section \ref{subsec::scatter_discussion}). We show the size-$z$ relations as a function of both redshift and lookback time as it is helpful to illustrate to the reader how galaxies grow as a function of linear cosmic time. Details of the studies that compose the individual points in Figure \ref{fig::redshiftandtime} can be found in Table \ref{tab::sizezsymbols}.

From Figure \ref{fig::redshiftandtime}, for a $10^{9.25}$\,\si{\solarmass} \ac{SF} galaxy, we predict $r_{e, \text{F444W}} = 0.76\pm0.46$\,\si{\kilo\parsec} from our F444W $r_{e}-M_{*}$ relationship (red point). The error associated with this size is the scatter about the relationship at this stellar mass. This $r_{e, \text{F444W}}$ agrees with a wide range of individual measurements in the literature at $z>4$ as well as the size-$z$ relationship from \citet{Sun2024} at $z\sim6$. There is also agreement with the lower-$z$ relationship from \citet{Martorano2024} extrapolated out to $z=6.1$. The exception to this is the relationship from \citet{vanderWel2014a} (grey dashed line), constrained between $z=0-1.5$. Extrapolating their size-$z$ relationship, \ac{SF} galaxy sizes are $\approx0.6$\,\si{\kilo\parsec} (factor $\approx1.8$) larger at $z=6.1$ than the other relationships we compare to. On the other hand, the relationship determined from the observations in \citet{Sun2024} agrees with our $r_{e, \text{F444W}}$ at $z=6.1$, although that relation over-predicts sizes at much lower redshifts (see also the relations in simulations from \citealp{Roper2022,Costantin2023}). The discrepancies between these studies can be explained by the fact their relationships are only measured at certain redshift ranges resulting in them failing to capture galaxy evolution at extrapolated redshifts that are not probed. All the simulated size-$z$ relationships that we compare to agree with our $r_{e}$ at $z=6.1$.

We fit a power-law of the form $\log_{10}(r_{e}\text{ / kpc}) = \beta\log_{10}(1 + z) + B$ to all the individual points we compare to (solid black line; see Table \ref{tab::sizezsymbols}), including our own $r_{e, \text{F444W}}$. The parameters of this fit are $\beta = -1.13\pm0.1$ and $B = 0.89\pm0.05$ which predicts $r_{e} \approx 7.8$\,\si{\kilo\parsec} at $z=0$. However, this fit and subsequently inferred $r_{e, \text{F444W}}$ may be biased by the ground-based observations of \ac{HAEs} by \citet{Stott2013a} and \citet{Paulino-Afonso2017} from HiZELS, which are less reliable than space-based measurements, and result in a significantly larger size estimate at $z=0$. Therefore, we also fit a power-law in the same form which excludes these points (dashed black line) with $\beta = -0.91\pm0.09$ and $B = 0.66\pm0.07$, predicting $r_{e} \approx 4.6$\,\si{\kilo\parsec} at $z=0$.

\subsection{Merger Fraction at \texorpdfstring{\boldmath$z$}{z} = 6.1}
\label{subsec::merger_rates}

\begin{figure*}
\centering

\includegraphics[width=1\columnwidth, trim=0.2cm 0 0 0.2]{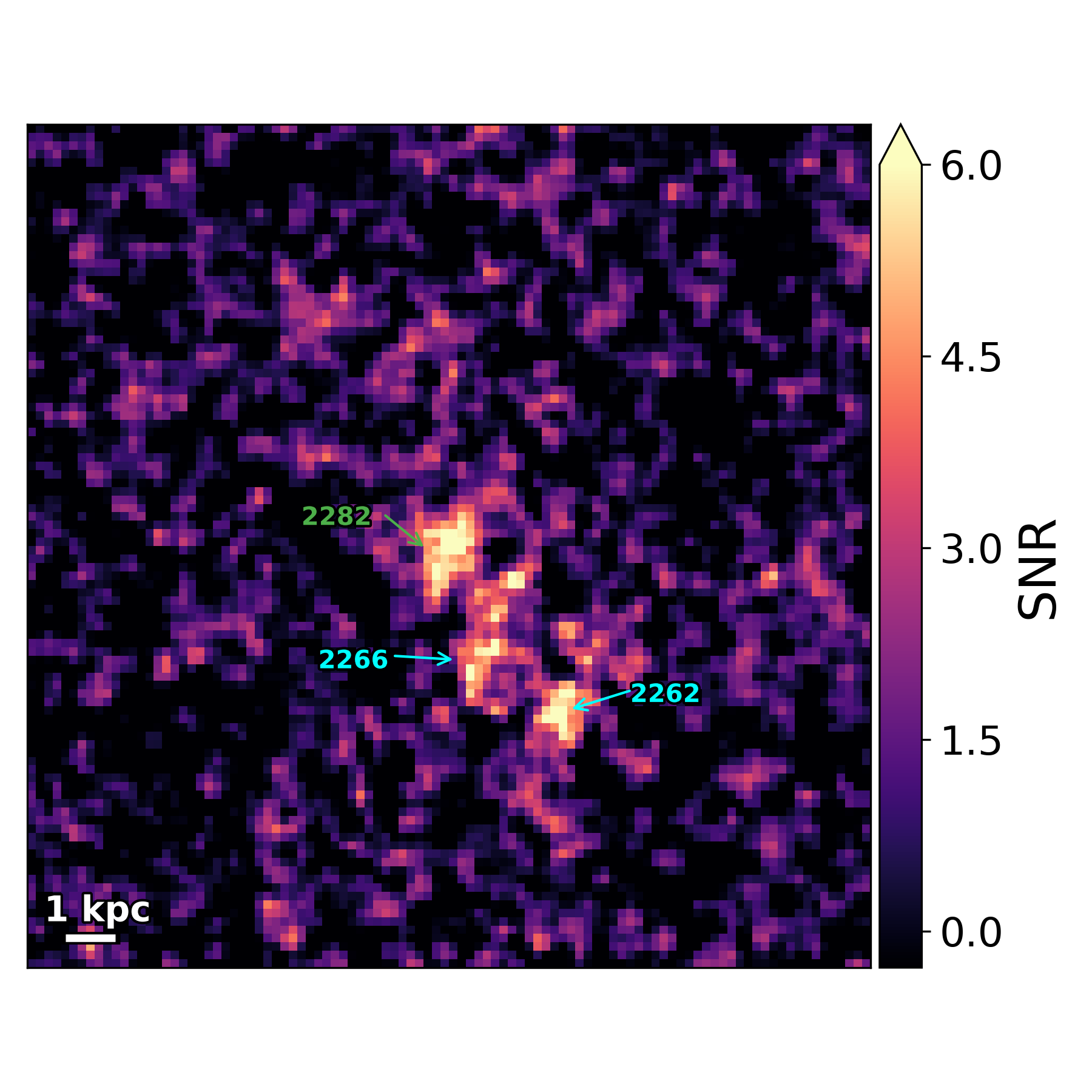}
\includegraphics[width=1\columnwidth, trim=0.2cm 0 0 0.2]{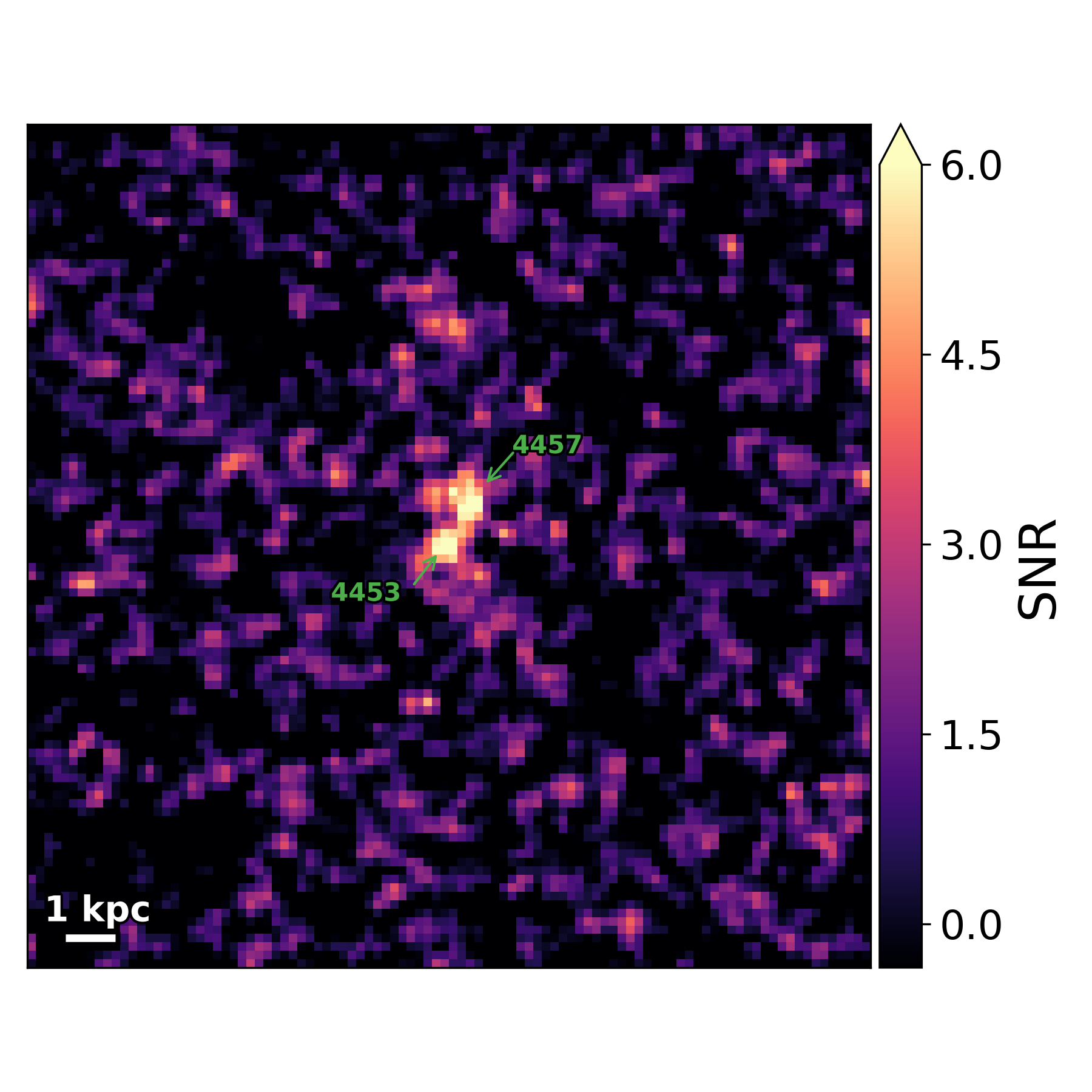}\vspace{-1cm}
\includegraphics[width=1\columnwidth, trim=0.2cm 0 0 0.2]{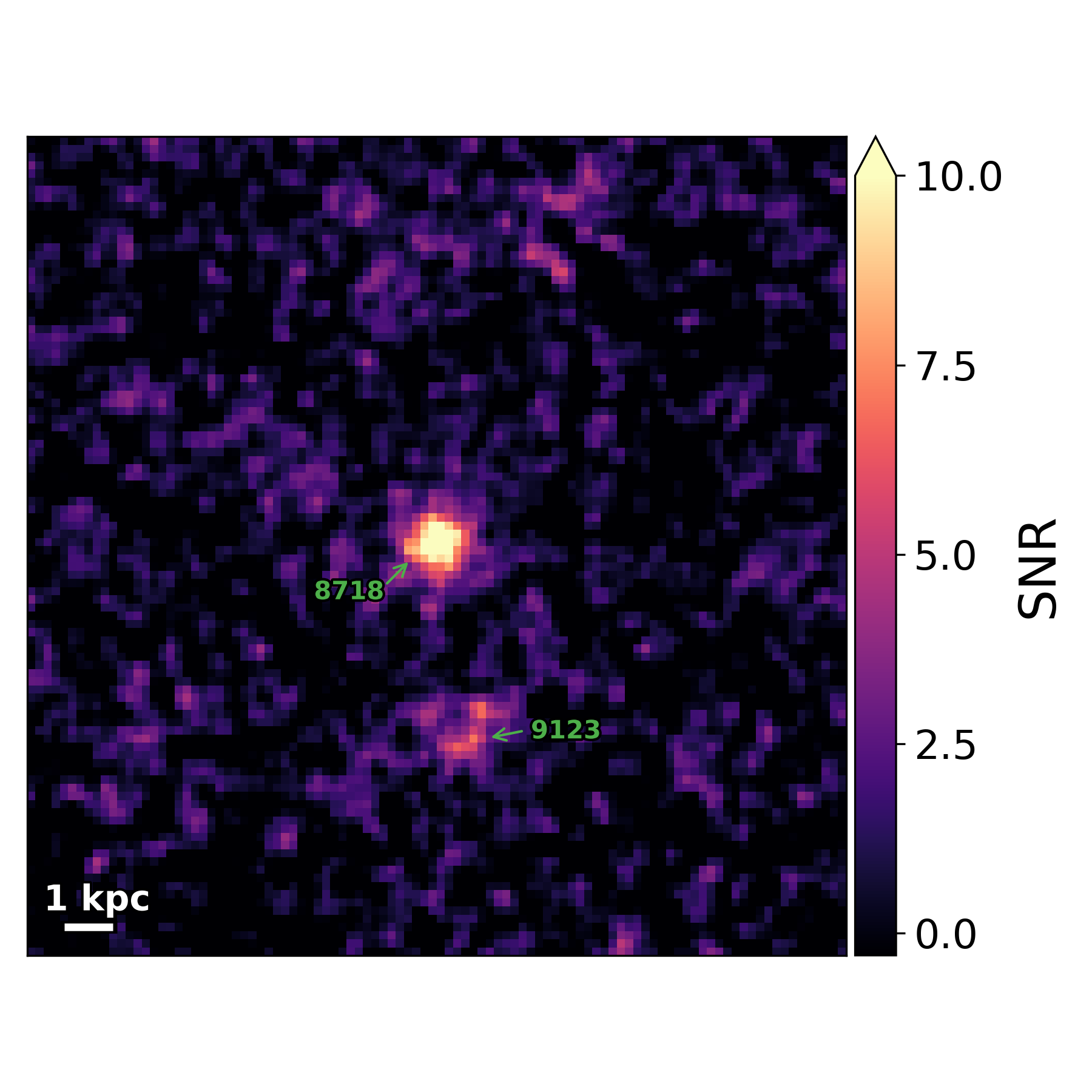}
\includegraphics[width=1\columnwidth, trim=0.2cm 0 0 0.2]{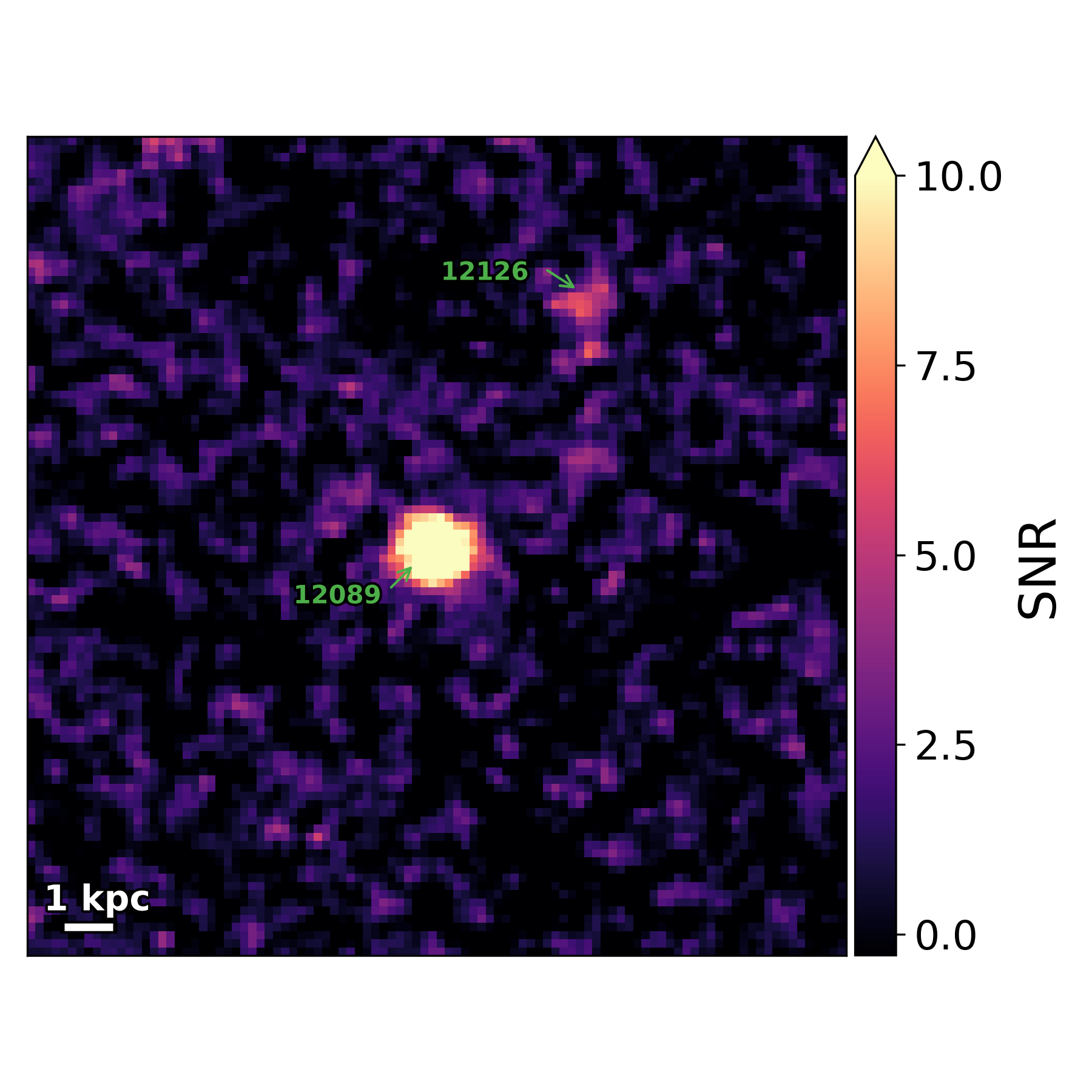}\vspace{-1cm}
\caption{$3\times3$\,arcsec$^{2}$ ($\sim17\times17$\,\si{\square\kilo\parsec}) cutouts of four merging systems in JELS F466N smoothed with a Gaussian kernel with \ac{FWHM} = 1.5 pixels. The colour bar indicates the signal-to-noise ratio (SNR) per pixel. These cutouts are centred on four different H$\alpha$-emitting galaxies at $z=6.1$ and are defined as a merging system from having at least one other NB-selected source within the cutout. Galaxies labelled in green are in the parent catalogue of \ac{HAEs} from \citet{Pirie2025a}. The cutout centred on galaxy 2282 (detected in F466N) has 2 other candidate sources within the cutout; galaxy 2266 has a similar photo-$z$ to galaxy 2282 ($z_{\text{phot}}=6.29$), but it has an excess significance parameter in F466N compared to F470N of $<2.5\sigma$ and $<3\sigma$ compared to F444W which are below the required threshold for a significant detection in F466N compared to either filter; galaxy 2262 has $>3\sigma$ excess compared to F444W, but its $z_{\text{phot}}=6.51$ is beyond the range required by \citet{Pirie2025a} to be in the catalogue of $5.5<z_{\text{phot}}<6.5$. These galaxies are therefore highlighted in blue to be clear that these are not in the parent catalogue of \ac{HAEs}.}
\label{fig::merging_systems}
\end{figure*}

As previously discussed, our final sample of 23 \ac{HAEs} at $z=6.1$ is drawn from the parent catalogue of 35 derived by \citet{Pirie2025a}. Present in their catalogue of sources - twelve of which we discard from our sample for reasons detailed in Section \ref{subsec::sampleselection} - are some systems with multiple \ac{HAEs}. Whilst we cannot accurately model the light profiles of those discarded galaxies, we can use them to approximate a merger fraction ($f_{\text{merger}}$). This is important to analyse because mergers are one of the primary mechanisms that contribute to galaxy growth \citep{Toomre1972}, including at the \ac{EoR} (e.g. \citealp{Dalmasso2024,Duan2025,Puskas2025,Westcott2025}; see Section \ref{subsec::mergerratecomparison}). To do this, we define a system as a merger if there is another nearby source with a similar photo-$z$ as detected by \citet{Pirie2025a} in their catalogue of PRIMER F356W-detected sources with $5.5<z_{\text{phot}}<6.5$. We use the parent catalogue of 35 \ac{HAEs} as the primary galaxies. We also imposed a confidence limit on photo-$z$ for the F356W detections such that the integrated redshift probability distribution function of photo-$z$ fitting, P($z$), is limited to detections with $\text{P}(z) > 0.7$. We then counted the number of detections that were within distances of $d\lesssim17$\,\si{\kilo\parsec} (corresponding to the width of our \texttt{GALFIT} cutouts), $d\lesssim25$\,\si{\kilo\parsec} and $d\lesssim50$\,\si{\kilo\parsec} to give a range of estimates for $f_{\text{merger}}$. This is a method for determining $f_{\text{merger}}$ known as \say{pair counting} (\citealp{Barnes1988}; see also \citealp{Patton1997,LeFevre2000,Bell2006a,Bell2006b}) defined as

\begin{equation}
f_{\text{merger}} = \frac{N_{\text{merger}}}{N_{\text{total}}},
\label{eq::merger_fraction}
\end{equation}

\vspace{5pt}
\noindent where $N_{\text{merger}}$ is the total number of pairs and $N_{\text{total}}$ is the total number of galaxies in our primary sample.

We find four systems that have multiple \ac{HAEs} with $z_{\text{phot}}\sim6.1$ detected within a $3\times3$\,arcsec$^{2}$ cutout, equivalent to $\sim17\times17$\,\si{\square\kilo\parsec}. We show these systems in Figure \ref{fig::merging_systems}. We smoothed these cutouts with a Gaussian kernel with \ac{FWHM} = 1.5 pixels to reduce some of the noise. We note that in the upper left panel of Figure \ref{fig::merging_systems}, galaxy 2266 has a $z_{\text{phot}}=6.29$ which is within the range required by \citet{Pirie2025a} of $5.5<z_{\text{phot}}<6.5$, but does not satisfy the other criteria to be in the parent catalogue. Specifically, it has an excess significance parameter in F466N compared to F470N of $<2.5\sigma$ and $<3\sigma$ compared to F444W which are below the required threshold for a significant detection in F466N compared to either filter. Moreover, while galaxy 2262 has $>3\sigma$ excess compared to F444W, its $z_{\text{phot}}=6.51$ is beyond the range required by \citet{Pirie2025a} to be in the catalogue. We therefore highlight the text for these galaxies in blue to be clear that these are not in the parent catalogue of \ac{HAEs}. However, given 2266 is a PRIMER F356W detection within the required redshift range, this is considered a merging $z\approx6.1$ system with 2282. This potential three-way merger could be an excellent candidate for follow-up with the \emph{JWST} NIRSpec Integral Field Unit \citep{Boker2022} or the Atacama Large Millimeter Array \citep{Wootten2009}.

Pair fractions in the literature are often selected based on the stellar mass ratio, $\mu$, of the pair. For \emph{major} mergers, this is typically defined as $\mu > 1/4$. Both the JELS parent catalogue of \ac{HAEs} and the PRIMER F356W-detected catalogue are complete down to $\approx10^{8.2}$\,\si{\solarmass}, so to calculate a major merger close-pair fraction, $f_{\text{maj. merger}}$, we set this as a lower mass limit for secondary galaxies, and $4\times10^{8.2} \approx 10^{8.8}$\,\si{\solarmass} for primary galaxies. For $f_{\text{maj. merger}}$, we also remove any object that exhibits point source activity as the stellar mass values from \ac{SED} fitting are biased to significantly higher masses. This mass cut, and the removal of point sources, reduces our primary galaxy sample to just 9 primary \ac{HAEs}.

\begin{table}
    
    \footnotesize
    \centering
    \caption{$f_{\text{merger}}$ from Equation \ref{eq::merger_fraction} for PRIMER F356W-detected sources within fixed distances, $d$, from systems in the parent catalogue of \ac{HAEs} in \citet{Pirie2025a}.}
    \label{tab::merger_fractions}
    \begin{tabular}{ccc} 
        \hline
        $d$ (\si{\kilo\parsec})              & $f_{\text{merger}}$   & $f_{\text{maj. merger}}$        \\
        \hline
$17$ (\ac{HAEs} only) $^{\ddagger}$  & $0.09\pm0.05$ & ...\\
$17$  & $0.29\pm0.09$ & $0.33\pm0.19$\\
$25$  & $0.43\pm0.11$ & $0.44\pm0.22$ \\        
$50$  & $0.71\pm0.14$ & $0.67\pm0.27$ \\        \hline{}
    \end{tabular}
\begin{flushleft}
$^{\ddagger}$ - only considering other \ac{HAEs} within the \texttt{GALFIT} cutouts.
\end{flushleft}
\vspace{-2mm}
\end{table}

We list our calculated close-pair fractions in Table \ref{tab::merger_fractions}. Within $d\lesssim17$\,\si{\kilo\parsec}, we find a $f_{\text{merger}} = 0.29 \pm 0.09$, which rises to $f_{\text{merger}} = 0.43 \pm 0.11$ ($f_{\text{merger}} = 0.71 \pm 0.14$) for detections within $d\lesssim25$\,\si{\kilo\parsec} ($d\lesssim50$\,\si{\kilo\parsec}). The systems in Figure \ref{fig::merging_systems} give us a merger fraction $f_{\text{merger}}\sim0.09$ if we only consider NB-detected sources within our cutouts. For $f_{\text{maj. merger}}$, our values are consistent with $f_{\text{merger}}$, though we find no examples of major mergers involving multiple \ac{HAEs} within $17$\,\si{\kilo\parsec}. We also note that our $f_{\text{maj. merger}}$ value is dominated by system 2282, which accounts for all pairs within $d\lesssim25$\,\si{\kilo\parsec}. We compare our merger fractions to values in the literature in Section \ref{subsec::mergerratecomparison}.


\section{Discussion}
\label{sec::discussion}


\subsection{Scatter of HAE Sizes at Low Stellar Mass}
\label{subsec::scatter_discussion}

From Figure \ref{fig::sizemassrelationship}, we see that at low stellar mass ($M_{*}<10^{8.4}$\,\si{\solarmass}), there is a larger scatter in $\log_{10}(r_{e}$) than at high mass for all filters. For example, the scatter of $\log_{10}(r_{e, \text{F444W}})$ in Figure \ref{subfig::bbsizemass} is $\sigma_{\text{scatter}} = 0.30$ dex, compared to $\sigma_{\text{scatter}} = 0.16$ dex at $M_{*} \geq 10^{8.4}$\,\si{\solarmass}. This increased scatter at low stellar mass may be a result of the \say{bursty} SFH of \ac{SF} galaxies at the \ac{EoR} which has been shown to have a greater impact on the evolution of less massive galaxies. Using the Feedback in Realistic Environments (FIRE; \citealp{Hopkins2014}) cosmological zoom-in hydrodynamical simulations, \citet{El-Badry2016} find that short-term stellar migration ($\sim100$\,\si{\mega\year}) can lead to significant fluctuations in $r_{e}$ by factors of $2-3$ during starbursts, and that this effect is strongest in low mass galaxies ($10^{7-9.6}$\,\si{\solarmass}; see also \citealp{Graus2019,Mercado2021}). Using the {\sc THESAN-ZOOM} simulations, \citet{McClymont2025b} showed that star formation in the early Universe is highly bursty on short ($\lesssim50$,Myr) timescales. Similarly, \citet{McClymont2025a} found that the size evolution of star-forming galaxies is strongly linked to starbursts, with galaxies alternating between phases of compaction and expansion which cause them to \say{oscillate} about the $r_{e}-M_{*}$ relationship. This rapid compaction arises because starbursts are typically centrally concentrated, before inside-out quenching subsequently increases their size once the burst subsides. Together, the results of \citet{McClymont2025a,McClymont2025b} suggest that \ac{EoR} galaxies undergo dramatic, short-term morphological transformations driven by bursty SFHs, potentially contributing to the scatter we observe in our $r_{e}-M_{*}$ relations. The SFH of our \ac{HAEs}, determined from \ac{SED} fitting by \citet{Pirie2025a}, is explored in greater detail in Section \ref{subsec::discussbbtonb}.

Given the assumption that the large scatter at low stellar masses may be caused by diverse SFH, as well as evidence in the literature from simulations, we looked for a connection between $r_{e, \text{F444W}}$ and the \ac{SF} properties of these $\lesssim10^{8.4}$\,\si{\solarmass} \ac{HAEs}. We did not find any correlation between $r_{e, \text{F444W}}$ and current \ac{SFR} measured from H$\alpha$, \ac{UV} continuum or \ac{SFR} derived from \ac{SED} fitting. Only a weak correlation is found between $r_{e, \text{F444W}}$ and the ratio of SED-fitted \ac{SFR} averaged over canonical timescales of 10\,\si{\mega\year} to SFR averaged over 100\,\si{\mega\year} (SFR$_{10}$/SFR$_{100}$), with a relationship in the form SFR$_{10}$/SFR$_{100}$ = ($0.58\pm0.65$)$r_{e, \text{F444W}} + (3.24\pm0.51)$, though we note the sample size for this fit is only 12 \ac{HAEs}. SFR$_{10}$/SFR$_{100}$ is a proxy of the burstiness of star formation (e.g. \citealp{Broussard2019}) and, as such, gives a good indication of the recent SFH of these galaxies, with \citet{Pirie2025a} showing that the \ac{HAEs} in the parent catalogue, particularly those at low stellar mass, are undergoing a recent upturn in star formation (see Section \ref{subsec::discussbbtonb} for further details). However, this analysis is only able to get an estimate of the time averaged SFR over $\leq100$\,\si{\mega\year}, which is the SFH from $z\approx6.7$. It is entirely possible that these galaxies may have undergone previous starbursts at $z\gtrsim6.7$ (see Section \ref{subsec::bbtonbsizesec}) which have contributed to the scatter we observe in the $r_{e}-M_{*}$ relationship at $z=6.1$.

Our analysis in Section \ref{subsubsec:galfit_uncertainties} shows that mock galaxies with magnitudes $\gtrsim27.5$\,mag and $r_{e} \lesssim 0.8$\,\si{\kilo\parsec} have overestimated $r_{e, \text{median}}$ recovered by \texttt{GALFIT} compared to $r_{e, \text{model}}$ (see Figure \ref{fig::galfiterrors}). While $r_{e, \text{median}}/r_{e, \text{model}}$ does not exceed a factor of 1.2 within the magnitude range of our sample, it is plausible that some of the less massive \ac{HAEs} (which tend to be fainter) are contributing to this scatter because they have elevated $r_{e}$ compared to their real size.

Additionally, as explored in Section \ref{subsubsec::nb_in_bb_emission}, we found some evidence that the F444W sizes of \ac{HAEs} at $<10^{8.4}$\,\si{\solarmass} are more impacted by the removal of H$\alpha$ contribution to the overlapping F444W, with $r_{e, \text{F444W}_{\text{sub}}}$ being larger than $r_{e, \text{F444W}}$ by factor of $1.10 \pm 0.32$ (Figure \ref{fig::nb_subtracted_size_ratios}). We also find that the scatter in $\log_{10}(r_{e, \text{F444W}_{\text{sub}}}$) reduces to $\sigma_{\text{scatter}} = 0.25$ dex. This suggests that the scatter in F444W at low stellar mass is being contributed to by H$\alpha$ emission in the BB, though this does not explain the scatter observed in other filters which are also $\gtrsim0.3$ dex. Ultimately, it is likely that a combination of all the reasons discussed above are contributing to the scatter of HAE sizes.


\subsection{Implications for Inside-Out Growth of Galaxies}
\label{subsec::discussbbtonb}

From Figure \ref{fig::bb_to_nb_size_ratio}, we find that the physical sizes of the stellar continuum of $z=6.1$ \ac{SF} galaxies are marginally bigger than sizes inferred from their H$\alpha$ emission. Given that the ratio for many of the individual objects is consistent with $\frac{r_{e, \text{F444W}}}{r_{e, \text{NB}}} = 1$, we can say that, at the very least, our results show that a significant stellar population has been built-up by the end of the \ac{EoR} that is comparable to the size of the \ac{SF} component and that this \ac{SF} region may be more centrally concentrated. It should be noted that using the H$\alpha$ emission as a direct proxy for the extent of the \ac{SF} component of galaxies can have its complications as a result of potential dust obscuration of the rest-frame optical light (e.g. \citealp{Wuyts2011,Nelson2012,Tacchella2015}). However, from their {\sc Bagpipes}\footnote{\url{https://bagpipes.readthedocs.io}} \citep{Carnall2018} SED fitting, \citet{Pirie2025a} have demonstrated that our sample of \ac{HAEs} are relatively dust-poor with a median $A_{V} = 0.23$. Indeed, rest-frame $R$-band light is less susceptible to dust attenuation compared to rest-\ac{UV} emission \citep{Calzetti2000,Salim2018}, which is often used in the literature to identify \ac{SF} regions (e.g. \citealp{Murphy2011,Mosleh2012,Ono2023,Morishita2024}) so any impacts on our overall sizes will not be as great as those studies. Additionally, we are observing the ratio of the stellar component to the \ac{SF} region at approximately the same wavelength between the NB and F444W filters ($\lambda \approx 4.4-4.7$\,\si{\micro\meter} rest-frame $R$-band). Therefore, should there be any significant impact from dust, it would affect the sizes in both the NB and F444W images approximately the same. A caveat to this is the possibility there may be a difference in the dust extinction for the stellar continuum and nebular components, but this remains uncertain at high-$z$ \citep{Sanders2025}.

Evidence in the literature suggests that galaxies experience what is known as \say{inside-out} growth \citep{vanDokkum2010}. In this paradigm, galaxies predominantly grow their mass and sizes from centrally concentrated \ac{SF} regions first before expanding out into, and indeed forming, an extended stellar disks towards lower redshifts. This transition of primary mass/size build-up from central regions to extended disks has been shown to come from either elevated \ac{SFR}s in the disk compared to the central bulge (e.g. \citealp{Dekel2014,Zolotov2015,Ellison2018,Shen2024a}) or from wet mergers (e.g. \citealp{Mihos1994a,Lin2008,Lapiner2023}). Whilst this is a reasonably well-known evolutionary track from Cosmic Noon, when the global \ac{SFR} peaks ($1\lesssim z \lesssim 3$; \citealp{Madau2014}), only recently has \emph{JWST} allowed inside-out growth to be observed directly, and in greater detail, out to the \ac{EoR}. For example, \citet{Baker2025} discovered a mature \ac{SF} galaxy at $z=7.43$ in the Great Observatories Origins Deep Survey-South (GOODS-S) field \citep{Giavalisco2004} from the \emph{JWST} Advanced Deep Extragalactic Survey (JADES; \citealp{Eisenstein2023}). From this, they were able to ascertain the recent and extended SFH of the galaxy which shows that the time-averaged \ac{SFR} over the prior 100\,\si{\mega\year} was highest in the central core of the galaxy, but over the most recent 10\,\si{\mega\year}, the \ac{SFR} is significantly higher in the disk, consistent with inside-out growth. Other studies have show similar consistencies with inside-out growth at $z\gtrsim6$ (e.g. \citealp{Morishita2024,Matharu2024,Kocevski2025}). More generally, prior to the launch of \emph{JWST}, observations found that rest-frame optical emission lines in galaxies at $z\gtrsim6$ had high rest-frame \ac{EW}s ($\gtrsim500$\,\si{\angstrom}; e.g. \citealp{Labbe2013,Smit2015,Roberts-Borsani2016,Endsley2020,Stefanon2022}), indicating strong \ac{sSFR}s at these redshifts which could imply rapid growth that aligns with the inside-out paradigm.

However, whilst evidence exists that galaxies evolve inside-out during the \ac{EoR}, it is becoming apparent that the SFH of galaxies at this epoch are complex and diverse. Galaxies have been shown to go through bursts of star formation (often referred to as bursty SFH; \citealp{Dressler2023,Dressler2024,Ciesla2024,Harshan2024a,Wang2024a,Looser2025}), which were previously predicted by simulations prior to \emph{JWST} \citep{Kimm2014,Ceverino2018,Furlanetto2022}. Recently, \citet{Endsley2025} analysed 368 $z\sim6$ Lyman-break galaxies in the GOODS fields and lensed fields surrounding the Abell2744 cluster and found a dramatic range of SFHs. They analyse the H$\alpha$ to \ac{UV} luminosity ratio ($L_{\text{H}\alpha}/L_{\text{\ac{UV}}}$) to infer the recent SFH of their galaxies and find their sample has properties consistent with extremely bursty SFHs, as well as finding that many of their galaxies have experienced strong recent \ac{SFR} upturns \emph{and} downturns. This followed a similar result from \citet{Endsley2024} who similarly concluded that $z\gtrsim6$ galaxies experienced bursty SFHs with evidence of strong recent downturns, this time using [OIII], H$\beta$ and H$\alpha$ \ac{EW}s.

For our sample of \ac{HAEs}, \citet{Pirie2025a} have demonstrated that they are going through a recent burst of star formation from their SED fitting, particularly those with stellar masses $M_{*}\lesssim10^{9}$\,\si{\solarmass}. As they point out, this was to be expected as H$\alpha$ emission is a good tracer of recent star formation ($\approx10$\,\si{\mega\year}; \citealp{Murphy2011}) compared to the \ac{UV}-continuum, for example, which can be used to probe longer timescales ($\gtrsim100$\,\si{\mega\year}; \citealp{Hao2011b}), though we caveat that \ac{UV} emission is produced by a combination of different stellar populations that range in lifespans ($\sim10-200$\,\si{\mega\year}), meaning it can only be confidently used to trace stellar populations older than $\gtrsim100$\,\si{\mega\year} in galaxies with steady-state star formation. However, in line with the studies we mention above, \citet{Pirie2025a} find that their results indicate that these \ac{HAEs} at $z=6.1$ have bursty SFHs.

\begin{figure*}
        \centering
        \includegraphics[width=\textwidth, trim=0 0 0 0, clip=true]{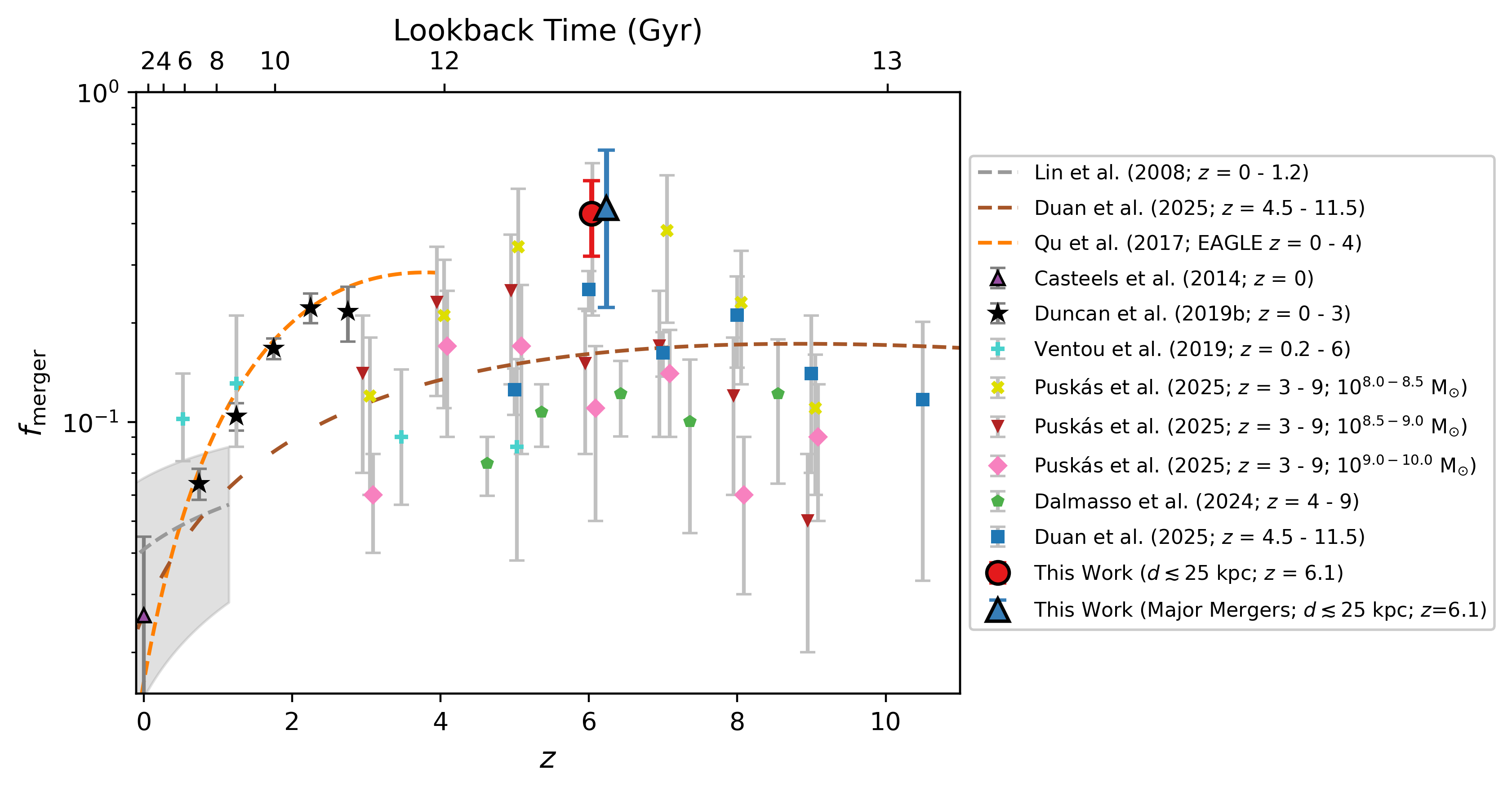}
    \caption{Our measured close-pair merger fraction, $f_{\text{merger}}$, based on PRIMER F356W detections at $z_{\text{phot}}\sim6$ within $d\lesssim25$\,\si{\kilo\parsec} of \ac{HAEs} in the parent catalogue of \citeauthor{Pirie2025a} (\citeyear{Pirie2025a}; large red circle). The large blue triangle represents our pair fraction for major mergers, $f_{\text{maj. merger}}$, with a stellar mass ratio of $\mu > 1/4$. Our $f_{\text{merger}}$ ($f_{\text{maj. merger}}$) value is offset by $+ (-) 0.1$ in redshift for clarity. The error on all individual points represent the standard error of the measured fraction at that redshift. Most comparisons in the literature use a similar close-pair fraction method to the one in this study, with the exception of \citet{Dalmasso2024} who use morphological statistical parameters for their fraction. The dashed lines from \citet{Qu2017} and \citet{Duan2025} show the redshift evolution of merger fractions as a power law with an additional exponential component of the form $f_{\text{merger}} = f_{0} \cdot (1+z)^{m} \cdot e^{\tau(a+z)}$. The grey dashed line indicates the power law evolution measured by \citet{Lin2008} at $z=0-1.2$ based on analysis of close-pair fractions. Their relationship is of the form $f_{\text{merger}} = f_{0} \cdot (1+z)^{m}$, with the grey shaded region indicating the $1\sigma$ error of their relationship. The brown dashed line is the observed evolution measured by \citet{Duan2025} between $z=4.5-11.5$, who use \citet{Casteels2014} as their $z=0$ point. Where the \citet{Duan2025} relationship becomes more spaced is where we extrapolate beyond their redshift range. The orange dashed line indicates the redshift evolution from the EAGLE cosmological hydrodynamical simulations \citep{Crain2015} as measured by \citet{Qu2017} from $z=0-4$.}
    \label{fig::mergercomparison}
\end{figure*}

In this context, given our results in this paper show that a significant stellar component has already been built up, and that the results of \citet{Pirie2025a} show that these same galaxies are currently undergoing a burst of star formation (especially those at $M_{*}\lesssim10^{9}$\,\si{\solarmass}), we conclude that the stellar component must have been built by some previous episodes of star formation and that these were likely bursts themselves. Moreover, we also believe that these episodes of star formation may have occurred at $z\gtrsim6.7$ ($\approx100$\,\si{\mega\year} prior to $z=6.1$) since the median SFR$_{10}$/SFR$_{100}\approx2.9$ for our sample of 23 \ac{HAEs}, suggesting the recent burst is significantly elevated compared to the averaged 100\,\si{\mega\year} SFR. These bursts of star formation may then be regulated by stellar feedback (i.e. \citealp{Ma2018,Katz2023,Shen2024b}) and/or mergers (see Section \ref{subsec::mergerratecomparison}). This is not necessarily in contradiction to inside-out growth evolution, as \citet{Pirie2025a} has shown that higher mass sources ($M_{*} \gtrsim 10^{9}$\,\si{\solarmass}) show evidence of a more consistent SFH. Additionally, the stochastic SFH exhibited by these sources may consistently be centrally concentrated given we are observing the current \ac{SF} region as being marginally smaller than the stellar component on average. This is best illustrated when we compare $r_{e, \text{F444W}_{\text{sub}}}$ to $r_{e, \text{NB}}$ and find that NB flux-subtracted F444W sizes are a factor of $1.26 \pm 0.14$ larger, which suggests a centrally concentrated \ac{SF} region. Stars formed during these bursts may then \say{fan} out with time as a result of stellar migration \citep{Schonrich2009,El-Badry2016} in a manner consistent with inside-out growth. Given the evidence of bursty star formation in the literature combined with our results, we suggest that a more complex approach to galaxy evolution is needed at the \ac{EoR}.


\subsection{Merger Fraction Comparisons}
\label{subsec::mergerratecomparison}

In this Section, we will compare our measured $f_{\text{merger}}$ and $f_{\text{maj. merger}}$ to values in the literature, from both observations and simulations. For the purposes of all comparisons, we will use our estimates at $5.5<z_{\text{phot}}<6.5$ within $d\lesssim25$\,\si{\kilo\parsec} (see Table \ref{tab::merger_fractions}) as this is the most comparable value to the comparison studies in Figure \ref{fig::mergercomparison}.

Figure \ref{fig::mergercomparison} shows how our close-pair fractions at $z=6.1$ compare to measurements in the literature for a range of redshifts. Most of these measurements use a similar close-pair fraction methodology to this work, with the exception of \citet{Dalmasso2024} (green pentagons) who use a combination of morphological statistical parameters, and we refer the reader to their paper for details on their methods. We highlight here that direct comparisons to their work should be noted with caution due to potential systematic effects that arise from the differences between our methods. We also note that the $z=0$ $f_{\text{merger}}$ from \citet{Casteels2014} (purple triangle) is used by \citet{Duan2025} as a supplementary zero-point for their $f_{\text{merger}}$-$z$ relationship (brown dashed line). All of the merger fractions at $z\gtrsim3$ in Figure \ref{fig::mergercomparison}, as well as the \citet{Casteels2014} value, are based on galaxy samples with a comparable stellar mass range to our own. The merger fractions of \citet{Duncan2019b} (black stars; typical stellar mass $M_{*} \sim 10^{10}$\,\si{\solarmass}), and the relation from \citet{Lin2008} (grey dashed line and shaded region; $M_{*} \sim 10^{10.7}$\,\si{\solarmass}) are based on stellar mass ranges that exceed our sample.

From Figure \ref{fig::mergercomparison}, we can see that our measured close-pair fractions of $f_{\text{merger}} = 0.43\pm0.11$ and $f_{\text{maj. merger}} = 0.44\pm0.22$ broadly agree with literature values at $z=6.1$ within uncertainty, though we find they are higher than the relationship of \citet{Duan2025} and the values from \citet{Dalmasso2024}. We particularly find excellent agreement with the $f_{\text{merger}}$ from \citet{Puskas2025}. Their results are derived from JADES observations of the GOODS-North (GOODS-N) and GOODS-S fields at $z\sim3-9$ for separations of $5<d<30$\,\si{\kilo\parsec}. As a result of their large sample size ($\sim$300 000), they split their $f_{\text{merger}}$ into different stellar mass ranges, and we find the best agreement with their $10^{8.0-8.5}$\,\si{\solarmass} values. This range also best matches the stellar mass of the parent catalogue of JELS \ac{HAEs} ($M_{*, \text{median}} = 10^{8.36}$\,\si{\solarmass}). Specifically, in this mass range, they find $f_{\text{merger}} = 0.41\pm0.20$ at $5.5<z<6.5$ which is consistent with all of our $f_{\text{merger}}$ and $f_{\text{maj. merger}}$ values in Table \ref{tab::merger_fractions}. 

Compared to the $z=0$ $f_{\text{merger}}$ from \citet{Casteels2014}, we find that close-pair fractions at $z=6.1$ are a factor of $\sim12$ higher than the local Universe for $d\lesssim25$\,\si{\kilo\parsec}. Comparing instead to the $z=0$ value inferred from the \citet{Lin2008} relationship, our close-pair fractions is a factor of $\approx8$ higher. Both of these comparisons demonstrate that the merger rate of galaxies during the \ac{EoR} is significantly higher than the local Universe, as also indicated by the other studies in Figure \ref{fig::mergercomparison}. However, there is evidence in the literature that the galaxy merger rates rise from $z=0$ before flattening and remaining consistent at $z\gtrsim3$, which is seen in both the observed $f_{\text{merger}}$-$z$ relationship in \citet{Duan2025} and results from the EAGLE simulations in \citet{Qu2017} (see Figure \ref{fig::mergercomparison}; see also \citealp{Conselice2014,Mundy2017,OLeary2021,Husko2022,Westcott2025}). Our results in Table \ref{tab::merger_fractions} for all separations, combined with results in the literature, suggest that galaxy mergers play an important role in galaxy evolution at the \ac{EoR}.

\section{Conclusions}
\label{sec::conclusions}

We utilised data from the \emph{JWST} Emission Line Survey (JELS; GO \#2321; PI: Philip Best; see \citealp{Duncan2025a,Pirie2025a}) to study the sizes of 23 \ac{HAEs} at $z=6.1$. Our sample is drawn from a parent catalogue of 35 \ac{HAEs} described in \citet{Pirie2025a}. We measured the size of both the ionised H$\alpha$ emission from $\lambda\sim4.7$\,\si{\micro\metre} NB data taken by JELS, and the stellar emission from $\lambda\sim4.4$\,\si{\micro\metre} PRIMER F444W images (both rest-$R$-band). In addition, sizes were also measured in PRIMER F277W (rest-\ac{NUV}) and PRIMER F356W (rest-$V$-band) to allow us to compare the light profiles of different stellar populations at the \ac{EoR}. We determine the sizes of galaxies from their half-light radii ($r_{e}$) which is measured using $n=1$ S\'ersic light profiles from \texttt{GALFIT}. We used these values to determine the size-mass ($r_{e}-M_{*}$) relationship of \ac{SF} galaxies at this epoch and compare to studies at lower redshift. We compared the average $r_{e, \text{F444W}}$ of our sample to a range of observational and simulated results in the literature from $z=0-11$. Using robust photo-$z$ detections in F356W at $z=6.1$, we were also able to determine an estimate of the merger fraction ($f_{\text{merger}}$) of galaxies during the \ac{EoR}. Our key results are summarised as follows:

\begin{enumerate}[label=\roman*), leftmargin=*]

    \item We observe a $r_{e}-M_{*}$ relationship in our sample of \ac{HAEs} in all observed NIRCam filters (Figure \ref{fig::sizemassrelationship} and Table \ref{tab::sizemassdata}). Our $r_{e}-M_{*}$ relationships are offset from those found at lower redshift. We find an offset of $-0.37\pm0.10$ dex in $\log_{10}(r_{e, \text{F444W}}\text{ / kpc})$ to the \citet{vanderWel2014a} relationship at $z=2.75$ for a fixed stellar mass of $10^{9.25}$\,\si{\solarmass}. This offset reflects a $\sim2.3-2.5$ factor increase in the sizes of the stellar component between $z=6.1$ and $z=2.75$ ($\approx1.4$\,\si{\giga\year}), suggesting \ac{SF} galaxies grow rapidly from the \ac{EoR} to Cosmic Noon.

    \item We measure the slope of the F444W $r_{e}-M_{*}$ relationship to be $\alpha_{\text{F444W}} = 0.14\pm0.12$ (Figure \ref{subfig::bbsizemass}). This slope is consistent with those found by \citet{vanderWel2014a} at both $z=0.25$ and $z=2.75$ as well as the $r_{e}-M_{*}$ slope of \ac{HAEs} at $z=0.4$ found by \citet{Stott2013a}. These results suggest that there is no significant redshift evolution in the slope of the $r_{e}-M_{*}$ relationship between $0.3\lesssim z \lesssim 6.1$.

    \item The average $r_{e, \text{F444W}}$ of a $10^{9.25}$\,\si{\solarmass} \ac{SF} galaxy at $z=6.1$, inferred from our $r_{e}-M_{*}$ relationship, is $0.76\pm0.46$\,\si{\kilo\parsec}. This value is in excellent agreement with a wide range of literature values at $z=6.1$, both from observations and simulations (Figure \ref{fig::redshiftandtime}).
    
    \item We measured the ratio of the F444W sizes to NB sizes for each of the galaxies in our sample (Figure \ref{fig::bb_to_nb_size_ratio}). This traces the size ratio of any established stellar component to the \ac{SF} region traced by ionised gas. We find that the median ratio of these sizes is $\frac{r_{e, \text{F444W}}}{r_{e, \text{NB}}} = 1.20\pm0.09$. Using rest-\ac{NUV} as a tracer of active star formation, we find $\frac{r_{e, \text{F444W}}}{r_{e, \text{F277W}}} = 1.14\pm0.07$ (Figure \ref{fig::bb_to_uv_size_ratio}). These measured ratios imply that \ac{SF} galaxies at $z=6.1$ have an already-established stellar component that is at least comparable to the size of the \ac{SF} region just $\sim900$\,\si{\mega\year} after the Big Bang. This also agrees with \ac{SF} galaxies exhibiting more centrally concentrated star formation at the \ac{EoR}.

    \item Previous analysis from \citet{Pirie2025a} indicates that these galaxies are undergoing a strong, recent starburst, with our sample of 23 \ac{HAEs} showing a median SFR$_{10}$/SFR$_{100}\approx2.9$. Given the evidence in the literature that galaxies at the \ac{EoR} have bursty SFH, we suggest that the established stellar component we observe in our sample may have resulted from episodes of star formation at $z\gtrsim6.7$ ($\gtrsim100$\,\si{\mega\year} prior to $z=6.1$). Additionally, we believe the large scatter ($\sigma_{\text{scatter}} \sim 0.30$) in the $r_{e}-M_{*}$ relationship at $M_{*} <10^{8.4}$\,\si{\solarmass} is being significantly contributed to by low mass galaxies being more affected by bursts of star formation giving them more diverse SFH. This could also be affected by \texttt{GALFIT} overestimating $r_{e}$ at faint magnitudes (Figure \ref{fig::galfiterrors}).

    \item We determine a close-pair fraction using close-pair counting based on PRIMER F356W $z_{\text{phot}}\sim6$ detections from \citet{Pirie2025a} and their parent HAE catalogue. We find $f_{\text{merger}}=0.43\pm0.11$ at $z=6.1$ using a galaxy separation of $d\lesssim25$\,\si{\kilo\parsec}. Using a stellar mass ratio of $\mu < 1/4$, we determine a close-pair fraction for major mergers of $f_{\text{maj. merger}} = 0.44\pm0.22$. These values agree with merger fractions in the literature at the \ac{EoR} (Figure \ref{fig::mergercomparison}). This shows mergers play an important role in galaxy growth from the \ac{EoR} to Cosmic Noon.

\end{enumerate}

\section*{Acknowledgements}

{This work makes use of {\sc AstroPy}\footnote{\url{http://www.astropy.org}}, a community-developed core Python package for Astronomy \citep{AstropyCollaboration2013,AstropyCollaboration2018,AstropyCollaboration2022}, as well as the {\sc NumPy} \citep{Harris2020} and {\sc SciPy} \citep{Virtanen2020} packages (see also \citealp{Oliphant2007}). All plots were created using the {\sc matplotlib} 2D graphics Python package \citep{Hunter2007}. Conversions between redshift and lookback time in our selected cosmological model were done using the Javascript cosmological calculator from \citet{Wright2006}\footnote{\url{https://astro.ucla.edu/~wright/CosmoCalc.html}}.

The authors would like to thank the anonymous referee for their constructive comments and suggestions which have strengthened the analysis of this work and improved the paper. The authors also gratefully acknowledge Ian Smail for providing valuable feedback and helping to guide the science of this paper. HMOS acknowledges support from an STFC PhD studentship and the Faculty of Science and Technology at Lancaster University. PNB is grateful for support from the UK STFC via grant ST/V000594/1 and  ST/Y000951/1. JSD acknowledges the support of the Royal Society via a Royal Society Research Professorship. LOF acknowledges funding by ANID BECAS/DOCTORADO NACIONAL 21220499. CLH acknowledges support from the Oxford Hintze Centre for Astrophysical Surveys which is funded through generous support from the Hintze family charity foundation. EI gratefully acknowledge financial support from ANID - MILENIO - NCN2024\_112 and ANID FONDECYT Regular 1221846.

For the purpose of open access, the authors have applied a Creative Commons attribution (CC BY) licence to any author-accepted manuscript version arising.}

\section*{Data availability}

The data underlying this article are available in the Mikulski Archives for Space Telescopes (MAST: \url{https://mast.stsci.ed}) Portal under proposal ID number 2321 (JELS imaging). Higher level data products, including all reduced mosaics in the JELS narrowband and broadband filters (v0.8 and v1.0) presented in \citet{Duncan2025a}, as well as associated catalogues presented in \citet{Pirie2025a}, are made available through the \href{https://datashare.ed.ac.uk}{Edinburgh DataShare} service. Any other data produced for the article will be shared on reasonable request to the corresponding author.


\bibliographystyle{mnras}
\bibliography{MNRAS_HMO_JELS_submission.bib}


\newpage
\appendix

\section{Free S\'ersic vs Fixed S\'ersic Sizes}
\label{sec::appendix}

\begin{figure*}
\centering
\begin{subfigure}{0.45\linewidth}
\includegraphics[width=\linewidth, trim=30 0 30 0]{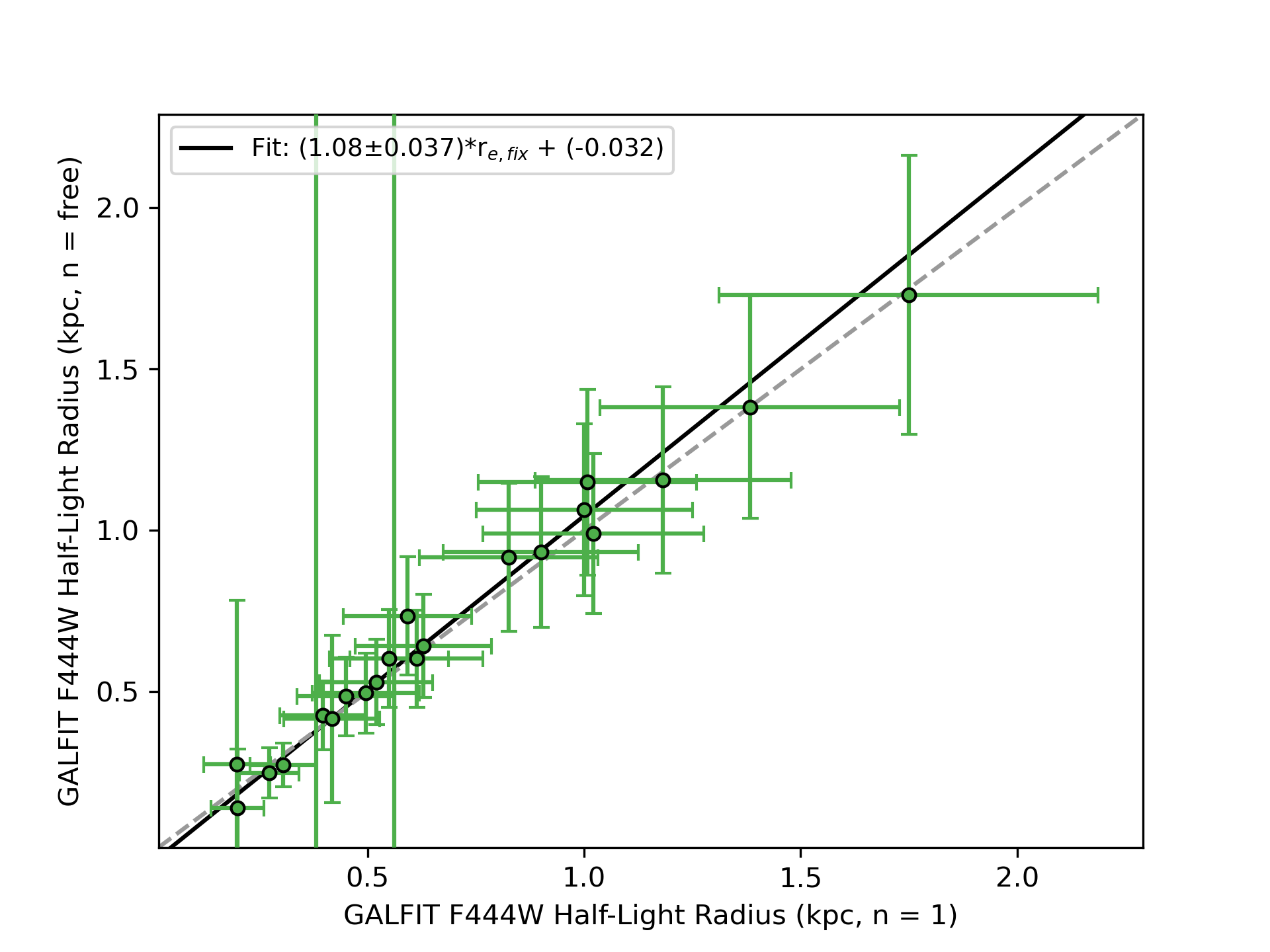}
\caption{}
\label{subfig::bbfreetofix}
\end{subfigure}
\hfill
\begin{subfigure}{0.45\linewidth}
\includegraphics[width=\linewidth, trim=30 0 30 0]{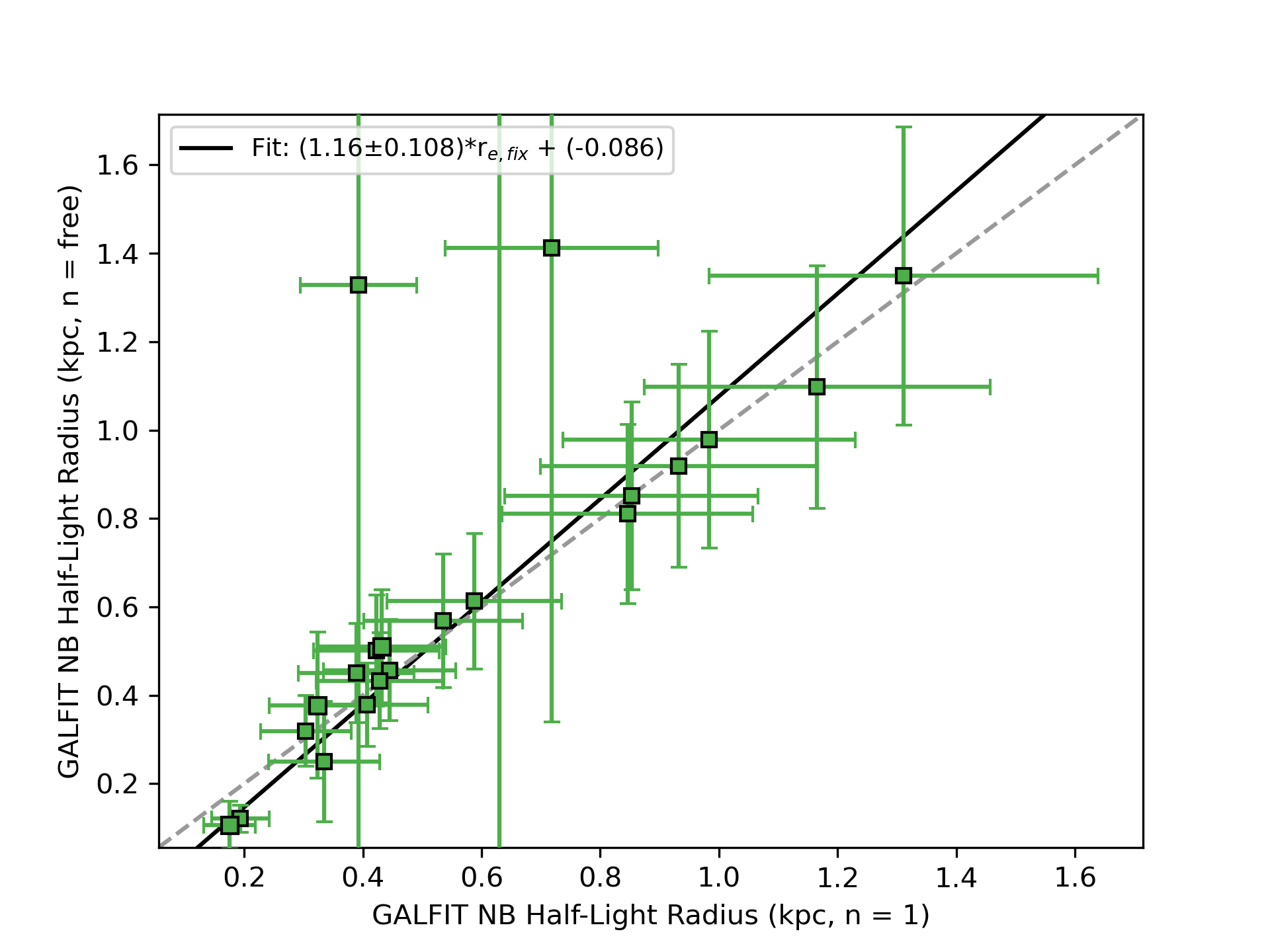}
\caption{}
\label{subfig::nbfreetofix}
\end{subfigure}

\caption{\emph{Left} - The measured free S\'ersic $r_{e}$ against fixed $n = 1$ S\'ersic $r_{e}$ in the F444W observations for each of the \ac{HAEs} in our sample at $z=6.1$. \emph{Right} - Same as the left panel, but for the NB observations. The grey dashed line indicates where the $r_{e}$ would be equal. The solid black lines show the best fit to the data.}
\label{fig::freevsfixedsizes}
\end{figure*}


\bsp    
\label{lastpage}
\end{document}